\documentclass[aps,pra,twocolumn,nofootinbib,10pt]{revtex4-2}

\usepackage{cancel}
\usepackage{footnotebackref}
\usepackage{xcolor}
\usepackage{footmisc}
\usepackage{subcaption}

\usepackage{ragged2e}
\usepackage{amsmath, amssymb, booktabs}
\usepackage{graphicx}
\usepackage{dcolumn}
\usepackage{bm}
\usepackage{physics}
\usepackage{hyperref}

\begin{document}

\preprint{APS/123-QED}

\title{Magnetic noise in macroscopic quantum spatial superposition } 

\author{Sneha Narasimha Moorthy$^{1, \,2}$}
\author{Andrew Geraci $^{3}$}
\author{Sougato Bose $^{4}$}
\author{Anupam Mazumdar$^{5}$}
\affiliation{ 
$^{1}$School of Physical Sciences, National Institute of Science Education and Research, Jatni 752050, India\\
$^{2}$Homi Bhabha National Institute, Training School Complex, Anushaktinagar, Mumbai 400094, India\\
$^{3}$Department of Physics and Astronomy, Northwestern University, 2145 Sheridan Road, Evanston, IL\\
$^{4}$Department of Physics and Astronomy, University College London, London WC1E 6BT, United Kingdom\\
$^{5}$ Van Swinderen Institute, University of Groningen, 9747 AG Groningen, The Netherlands\\}

\date{April 17, 2025}

\begin{abstract}
{ In this paper, we will show how random fluctuations in the 
magnetic field gradient will jitter the paths of a matter-wave interferometer randomly, hence, decohere the quantum superposition. 
To create a large spatial superposition with nanoparticles, we envisage embedding a spin in a nanoparticle as a defect and applying an inhomogeneous magnetic field as in a Stern-Gerlach-type experiment to create a macroscopic quantum superposition. Such matter-wave interferometers are the cornerstone for many new fundamental advancements in physics; particularly, adjacent matter-wave interferometers can use entanglement features to test physics beyond the Standard Model, test the equivalence principle, improve quantum sensors, and test the quantum nature of spacetime in a lab. In particular, we will study how white and flicker noise induces decoherence for a setup involving superconducting wires embedded in a chip. We will show that to obtain a tiny spatial superposition of a nanometer separation, $\Delta x \sim {\cal O}(10^{-9})$m and to minimize decoherence, $\Gamma\leq \order{\frac{\omega_0}{2\pi}}$, where $\Gamma$ is the decoherence and $\omega_0$ is the frequency of the oscillator, we will need current fluctuations to be $\delta I/I\leq {\cal O}(10^{-8})$, which is not impossible to obtain in superconducting wire arrangements. For such tiny fluctuations, we demonstrate that the Humpty-Dumpty problem in a matter-wave interferometer arising from a mismatch in position and momentum does not cause a loss in contrast.
}
\end{abstract}

\maketitle


\section{Introduction}

The color-center defects, known as a nitrogen-vacancy (NV)-centre in a nanodiamond have a multitude of applications from quantum metrology to quantum sensors~\cite{Doherty_2013}. Furthermore, matter-wave interferometers with an NV-centered nanodiamond open up a unique testing ground for fundamental physics, such as testing the quantum nature of spacetime.
Recently, the authors of \cite{Bose:2017nin,ICTS,Marshman:2019sne,Bose:2022uxe}, see also  \cite{Marletto:2017kzi} proposed a protocol to test the quantum nature of spacetime in a lab via spin entanglement witness~\cite{Bose:2017nin}, see also~\cite{Danielson:2021egj,Carney_2019,Carney23_nu,Biswas:2022qto,christodoulou2023locally,christodoulou2019possibility,Rufo:2024ulr,Hanif:2023fto}. Entanglement provides a bonafide quantum correlation, which cannot be mimicked by any classical feature, see~\cite{Horodecki:2009zz}. If two masses in quantum superpositions can be entangled {\it solely} via gravity, then the spacetime ought to behave like a quantum entity~\cite{Marshman:2019sne,Bose:2022uxe}, known as the QGEM (Quantum Gravity Mediated Entanglement of Masses) protocol.
It is well-known that entanglement between two quantum systems requires quantum interaction, or quantum mediator, which is the essence of a theorem known as local operations and classical communication (LOCC), which cannot entangle the two quantum systems.

Entanglement based protocols can be used to witness relativistic corrections to the Coulomb potential~\cite{Toros:2024ozf}, post-Newtonian corrections to low energy quantum theory of gravity~\cite{Toros:2024ozu}, massive graviton in the context of brane-world scenarios~\cite{elahi2023probing},  modified theories of low energy quantum gravity~\cite{Vinckers:2023grv,chakraborty2023distinguishing}, physics beyond the Standard Model~\cite{Barker:2022mdz}, test of quantum version of the equivalence principle~\cite{Bose:2022czr}. Most importantly, we can also test the entanglement between matter and photon degrees of freedom~\cite{Biswas:2022qto}, which will certify the spin-2 nature of the graviton as a mediator, and will provide a quantum entanglement version of the light-bending experiment due to quantum natured graviton. Also, matter-wave interferometers can ideally act as a quantum sensor, detecting external accelerations due to gravity~\cite{Wu:2022rdv,Wu:2024bzd,Toros:2020dbf}, electromagnetic interactions~\cite{Schut:2023tce,Fragolino:2023agd,Zhang:2025fxs,Sinha:2022snc,Zhou:2025jki}, and high-frequency gravitational waves~\cite{Marshman:2018upe}.

Any of these experimental protocols will be subject to decoherence~\cite{bassireview}  due to many external interactions which are random. In general, any matter-wave interferometer is sensitive to external noise and fluctuations in ambient pressure, temperature, current, voltage, etc. ~\cite{Toros:2020dbf,Rijavec:2020qxd,
vandeKamp:2020rqh,Schut:2021svd,Schut:2023eux,Schut:2023hsy,Fragolino:2023agd,Schut:2024lgp}. There are phonon-induced noise~\cite{Henkel:2021wmj,Henkel:2023tqe,Xiang:2024zol}, and fluctuation in the spin degrees of freedom during the dynamics of rotation of the rigid body~\cite{Japha:2022phw,Zhou:2024pdl}, all leading to dephasing and decoherence, see~\cite{bassireview,ORI11_GM,Hornberger_2012}, and loss of contrast~\cite{Englert,Schwinger,Scully,Margalit:2020qcy}
 In the backdrop of all the above extensive studies of decoherence, however, the most important is arguably the systematic noise stemming from the nature of the protocol itself. As a Stern-Gerlach mechanism is initiated by magnetic field gradients, these gradients form the core of the systematic noise. How much of such a noise can be tolerated then becomes an important question intrinsic to the process of wavefunction splitting and recombination.

For NV-centered nanodiamonds, it is possible to create a spin superposition of 
$|+1\rangle$ and $|-1\rangle$ states, and then by applying the Stern-Gerlach force on the spin, which is susceptible to the external inhomogeneous magnetic field~\cite{Margalit:2020qcy,amit2019t}, it is possible to separate the centre of mass motion of the nanodiamond onto left and right trajectories, before bringing them together to perform one-loop interferometer, thereby creating a Schr\"odinger cat state. There are many variants of this scheme, see~\cite{Wan16_GM,Scala13_GM,
Bose:2017nin,Pedernales:2020nmf,Marshman:2021wyk,Marshman:2018upe,Zhou:2022epb,Zhou:2022frl,Zhou:2022jug,Zhou:2024voj,Braccini:2024fey,Rizaldy:2024viw,Zhou:2024pdl,japha2022role,japha2021unified}. One might expect that we would be able to cool the initial state of the center-of-mass motion~\cite{Deli__2020,Piotrowski_2023,Kamba:2023zoq,Bykov:2022xji,Perdriat:2024xiy} before creating the spatial superposition. 

In the QGEM protocol, precise control over the external noise sources is crucial to maintain coherence in the spatial superposition states. Since, the setup prescribed in \cite{Bose:2017nin} is based on diamagnetic levitation, see~\cite{Elahi:2024dbb,Schut:2023eux,Schut:2023hsy}, and to create the superposition also we rely on the inhomogeneous magnetic field. Hence, any fluctuations in the magnetic field will give rise to the random jitter, decoherence. Typically these random jitters will be the dominant source of dephasing. The fluctuations in the magnetic field will originate from the current-carrying wires embedded in the chip, see~\cite{Elahi:2024dbb}.

The current analysis focuses on characterizing the dephasing introduced by such noise by modelling it using classical stochastic processes at the Lagrangian level. We specifically consider two types of noise: Gaussian white noise and \( 1/f \) (flicker/pink) noise, and investigate their respective impacts on the phase evolution of the interferometric superposition. We will then analyze the Humpty-Dumpty problem in the one-loop matter-wave interferometer, see~\cite{Englert,Schwinger,Scully,Margalit:2020qcy}. The issue is that the fluctuations in the path will affect the position and the momentum of both the left and right arms of the interferometer, hence, any classical mismatch will turn into a loss of contrast in the interference pattern.  We will consider the constraints on the current fluctuations from the dephasing and show how much is the mismatch in the classical trajectories in the matter-wave interferometer.

\section{Fluctuations and dephasing}\label{sec.fluctuations and dephasing}

Let us consider the following general Lagrangian for an interferometer:
\begin{align}
    L_j = &\frac{1}{2}mv_j^2 - A_jx_j^2 - B_jx_j - C_j \quad \forall j\in{R,L} \label{eq.Lag_j_gen}
\end{align}
where, $j=R,~L$ denote the two arms of the interferometer.
The coefficients $A_j$, $B_j$ and $C_j$ are system-dependent constants in an ideal noiseless scenario. However, in a realistic experimental setup, these parameters acquire time-dependent fluctuations due to noise sources in the system. The ensemble statistics of these fluctuations depend on the nature of the noise present. Here $m$ is the mass of the interferometer, and $v_j$ are the velocities of the two arms. We will show how linear in position and quadratic terms appear in the Lagrangian by taking a simple example of the Stern-Gerlach type interferometery~\cite{amit2019t,PhysRevLett.123.083601,Pedernales:2020nmf,Marshman:2021wyk,doi:10.1126/sciadv.abg2879}.

The phase difference between the two paths, $\Delta \phi(t)$ (corresponding to the two interferometer paths) accumulates throughout the experiment. 
\begin{equation}
    \Delta \phi = \int_{t_i}^{t_f} (L_R - L_L)\,dt
\end{equation}
In the absence of noise, this phase difference purely reflects the differential evolution of the two paths. However, noise in the system introduces additional stochastic contributions $\delta \phi$ to this phase difference, leading to dephasing.

Note that, if $C_j$ doesn't contain any element that introduces fluctuations, it does not affect $\delta \phi$. We shall see that this will be the case in our setup, see below. The equations of motion change randomly if there is an external noise. Let $x_j$ be the trajectory that satisfies the equations of motion (EOM) before the introduction of the noise, and $x_j^{tot}$ be the trajectory that satisfies the EOM after the introduction of noise. Then, 
\begin{equation}
    \delta x_j = x_j^{tot} - x_j \label{eq.def_dev_trajd}
\end{equation}
is the deviation in trajectory due to the introduction of noise. However, in the first analysis, we shall always assume that the two trajectories meet to interfere after the completion of one loop. Later on, we will relax this condition and will study the Humpty-Dumpty problem which will lead to the loss in contrast.

At the lowest order, we assume that 
the fluctuations affect the coefficients $A_j,~B_j,~C_j$ only, and not the trajectories themselves.
Hence, we obtain:
\begin{align}
    \delta \phi = &\int_{t_i}^{t_f} \bigg[(\delta A_Lx_L^2-\delta A_Rx_R^2) \nonumber\\
    &+ (\delta B_L x_L-\delta B_R x_R) + (\delta C_L-\delta C_R)\bigg]\,dt \label{eq.genphasedif}
\end{align}
%
%
Since noise is a stochastic process, we can only characterize ensemble averages of noise. Hence, we can theoretically calculate the average deviation in phase difference that the noise contributes to. If the experiment lasts for a total time \(T_{exp}\), then any noise with a frequency lower than ${T^{-1}_{exp}}$ (i.e., time period longer than \(T_{exp}\)) would be effectively static over the course of the experiment. This means it would appear as a constant offset rather than as a fluctuating noise. Frequencies just below ${T^{-1}_{exp}}$ will start to be averaged out, and ones much below are effectively static. In contrast, for noise to have a measurable effect as a fluctuating signal within the experiment, its frequency should be at least comparable to ${T^{-1}_{exp}}$ or higher. Thus, we set the lower frequency cutoff as \( \omega_{min} = {2\pi}/{T_{exp}} \),  ensuring that all relevant frequencies contributing to the phase decoherence are considered. To quantify the impact of the noise in the phase evolution, we evaluate the ensemble-averaged variance of the phase difference, denoted as $\Gamma$, assuming that the noise is Gaussian:
\begin{align} 
    \Gamma =&\, \lim_{\tau\to\infty}\frac{1}{\tau}E[(\delta\phi(\tau))^2] \label{eq.Gammaandensemble}\\
    =& \,\frac{1}{\hbar^2} \int_{\omega_{min}}^{\infty} \ S(\omega) \bigg|\Sigma_i\sqrt{F_i(\omega)}\bigg|^2 \,d\omega\label{eq.Dephasing_genform}
\end{align}
where $\tau$ is the largest time period of the contributing noise~\footnote{The contribution of noise to dephasing can be regularized either by constraining the integration range in frequency space (i.e., imposing a lower bound \( \omega_{\min} \sim {2\pi}/{T_{exp}}\)) or by limiting the noise time-period \( \tau \sim T_{exp} \). While doing both is mathematically rigorous, constraining either variable is typically sufficient to isolate the relevant spectral behaviour of the noise.
} and $\omega$ is the angular frequency of noise, \( S(\omega) \) is the power spectral density (PSD) of the noise, and \( F_i(\omega) \) are a system-dependent response/transfer function that encodes how these fluctuations influence the accumulated phase. Eq.(\ref{eq.Dephasing_genform}) is a result of the Weiner-Khinchin Theorem, see~\cite{Khintchine1934,Wiener:1930}. We refer to Appendix.\ref{appendix:A} for the derivation of eq.(\ref{eq.Dephasing_genform}).

The function \( F(\omega) \) contains information about how susceptible the system is to noise at a given frequency. In the initial analysis we neglect the effect of the noise on the trajectory itself so that we may assume that at the end of a one-loop interferometer, the paths meet. Then we compute the transfer function by considering the difference in the trajectory between the right and the left arms of the interferometer in the frequency domain. They are given by the linear and the quadratic part of the EOM, as:
\begin{align}
    F_1(\omega) &\propto \left| \int_{t_i}^{t_f} dt\, (x_R - x_L) e^{i \omega t} \right|^2\\ \label{eq.F1}
    F_2(\omega) &\propto \left| \int_{t_i}^{t_f} dt\, (x_R^2 - x_L^2) e^{i \omega t} \right|^2 
\end{align}
In case the sign of a term in the Lagrangian contributing to the noise is dependent on the spin of the state (eg. $\delta B_R = -\delta B_L$ in eq.(\ref{eq.Lag_j_gen})), then the following transfer functions would contribute as well: 
\begin{align}
    F_3(\omega) &\propto \left| \int_{t_i}^{t_f} dt\, (x_R + x_L) e^{i \omega t} \right|^2 \\
    F_4(\omega) &\propto \left| \int_{t_i}^{t_f} dt\, (x_R^2 + x_L^2) e^{i \omega t} \right|^2 \label{eq.F4}
\end{align}
In eqs.(\ref{eq.F1}-\ref{eq.F4}), we have assumed that $\delta A_L$ and $\delta B_L$ are some linear functions in $\delta A_R$ and $\delta B_R$, respectively. 

The noise spectrum \( S(\omega) \) depends on the statistical properties of the fluctuations/noise. For illustration, we will consider two kinds of noise, one is white noise to get an estimation of how large is the dephasing we can tolerate; and the second, flicker noise as superconducting wires are known to exhibit 1/f magnetic noise due to surface spin fluctuations and disorder—making it a potentially dominant source of low-frequency decoherence in our setup. This type of noise is very common in magnetic fields generated from superconducting wires, see for a review~\cite{Sendelbach2013,sergeenkov1999,kelly2024}. We discuss their properties below:
\begin{enumerate}
    \item For white noise, \( S(\omega) = A^2 \) is constant across all frequencies. White noise is characterized by the following statistics:
\begin{align}
    E[\delta\eta(t)] &= 0  \label{eq.whitenoise_statprop1}\\
    E[\delta\eta(t)\delta\eta(t')] &= A^2\delta(t - t')\label{eq.whitenoise_statprop2}
\end{align}
From eq.(\ref{eq.whitenoise_statprop2}), we obtain the following PSD for white noise {~\footnotetext{$\delta\tilde{\eta}_\tau(\omega) = \frac{1}{2\pi}\int_{-\tau}^{\tau}{\delta\eta}(t)\ e^{i\omega t}\,dt$\label{fn_S}}.}
\begin{align}
    S_{\eta\eta}(\omega) &=  \lim_{\tau \to \infty}\frac{1}{\tau}E[\delta\tilde{\eta}_\tau(\omega)\delta\tilde{\eta}_\tau^*(\omega)]  =A^2 \label{eq.gen_stat_noise} 
\end{align} 

Here, \( A \) is a constant that depends on the characteristics of the noise source. In Section~\ref{sec.4}, we constrain \( A \) based on the maximum tolerable dephasing in the simplest matter-wave interferometer.

    \item For flicker noise, \( S(\omega) \propto 1/\abs{\omega}^\alpha \), which leads to stronger low-frequency contributions, where $\alpha \in [0.5,1.5]$~\footnote{$\alpha = 0$ gives back white noise statistics.$\alpha = 2$ gives back Brownian noise statistics.}, see~\cite{PhysRevB.32.736, RevModPhys.53.497}. The magnetic field is considered to be produced by current flowing in a conductor or a superconducting wire. Any noise in the current will cause a noise in the magnetic field gradient. We consider the flicker noise contribution with the following PSD~\cite{PhysRevB.32.736, RevModPhys.53.497}:
\begin{align}
    S_{II}(\omega) &= \lim_{\tau \to \infty}\int_{-\frac{\tau}{2}}^{\frac{\tau}{2}} E[I(t)I(t')] e^{i\omega (t-t')} \,dt\,dt' \nonumber\\
    &=  E[\tilde I(\omega)\tilde I^*(\omega)] 
= \frac{KI^2}{|\omega|^\alpha}
\end{align}
where, $I$, is the DC current without fluctuations and $K$ is a source dependent constant~\footnote{The universal flux noise for SQUIDS and superconducting thin films is given by:\(
S_{\Phi} = {A}/{f^{\alpha}}
\)
with the magnitude of \( A \sim 5 - 10 \, \mu\Phi_0 / \sqrt{\text{Hz}} \) at 1 Hz, and \( 0.58 < \alpha < 0.80 \) \cite{Sendelbach2013}. Where $S_{\Phi} = {S_{\eta\eta}}/{(A^2L^2)}$, L and A are the length and area of the superconductor. 
}. $\alpha\approx 1$~\footnote{ For elements, Al: $\alpha\in [1,1.1]$ and Nb: $\alpha\in [1,1.4]$ \cite{PhysRevB.32.736}} in case of superconducting wires made up of Nb. 

Since we are levitating the nanoparticle at a distance $d$ away from a current carrying chip, see~\cite{Elahi:2024dbb}, we will consider a thin wire limit and that the wire length is much larger than the nanoparticle, such that we can take the infinite wire limit in our computations~\footnote{The current in eq.\ref{eq.MF} is an effective current from a combination of wires, which is responsible for creating the superposition. Note that current-carrying wires on the chip are also required to levitate the nanodiamond, see~\cite{Elahi:2024dbb}, and  \cite{folman2008wires}. However, here we are interested in the wire configuration which will trigger the linear magnetic field gradient 
along one spatial dimension to create a superposition, see~\cite{Pedernales20_GM,Marshman:2021wyk,marshman2024entanglement}.}.
\begin{equation}
    B = \frac{\mu_0 I}{2\pi d} \label{eq.MF}
\end{equation}
The magnetic field gradient is given by:
\begin{equation}
    \eta_0 = \frac{\partial B}{\partial x} = -\frac{\mu_0 I}{2\pi d^2}\label{eq.eta0_exp}
\end{equation}
Hence, the PSD of the noise due to the gradient of the magnetic field is given by~\cite{PhysRevB.32.736, RevModPhys.53.497}:
\begin{equation}
    S_{\eta\eta}(\omega) =\frac{\mu_0 S_{II}}{2\pi d^2} = \frac{\mu_0 KI^2}{ 2\pi d^2 |\omega|^\alpha}\label{eq.PSD_flicker}
\end{equation}
\end{enumerate}
By analyzing the dependence of \( \Gamma \) on the noise spectral density, we aim to quantify the impact of the two types of magnetic field fluctuations on the coherence of spin superpositions in a Stern–Gerlach-type interferometer~\cite{amit2019t,doi:10.1126/sciadv.abg2879,Pedernales20_GM,Marshman:2021wyk}. In particular, we consider the parameter regime relevant to the experimental proposal of levitating the nanodiamond, see~\cite{Elahi:2024dbb}, and derive constraints on the associated experimental parameters.

\section{NV-centered nanodiamond}

For illustration, consider an NV-centered nanoparticle with a mass $m$ in a spin superposition entering a Stern-Gerlach-type interferometer at $t=0$. The following is the Lagrangian for each of the states composing the superposition~\cite{Pedernales:2020nmf,
Marshman:2021wyk,Zhou:2022epb}~\footnote{We are assuming that the nanoparticle is levitated via diamagnetic levitation. We are ignoring the effects of gravity or any fluctuations due to gravity. Such fluctuations to some extent were discussed in our earlier papers, see~\cite{Toros:2020dbf,Wu:2022rdv}.}:
\begin{align}
    L_j = &\frac{1}{2}mv_j^2 - \frac{1}{2}m\omega_0^2x_j^2 - \big(S_{xj}\hbar\gamma_e\eta_0 - \frac{ \chi_\rho m}{\mu_0}B_0\eta_0\big)x_j\nonumber\\
    &+\frac{\chi_\rho m}{2\mu_0}B_0^2 - S_{xj}\hbar\gamma_eB_0 - \hbar DS_z^2 \quad \forall j\in{R,L} \label{eq.Lag_j}
\end{align}
where the external magnetic field is taken to be $\mathbf{B} = (B_0+\eta_0 x)\hat{x}$~\footnote{We assume that the magnetic field is in the $x-y$ plane, therefore, $B_y=-\eta_0 y \hat y$. However, we are assuming that the superposition will take place in one dimension. We are assuming an ideal case where we take the initial condition of $y=0$. In reality, it will be extremely hard, and this will require knowing the centre-of-mass motion along $x,z$ directions extremely well. We will need to initiate the experiment at $y=0$, in which case there will be no displacement due to the external inhomogeneous magnetic field along this direction. 
}, and $\chi_{\rho}= -6.286\times10^{-9} \text{m}^3 \text{kg}^{-1}$ (for nanodiamond) represents the mass magnetic susceptibility of the particle. We consider dynamics only along the x-axis. $S_{xj}$ is the spin of the particle in state j. $R, L$ represent the two spin (in the x-basis) and spatial superposition states ($S_{xR} = 1$ and $S_{xL}=-1$); they denote the two arms of the interferometer. Note that the spin and the center of mass are entangled in an SG interferometer. $D$ in the last term refers to the zero-field splitting which in the case of NV centres is $D=2.87GHz$\cite{Gruber2012-1997} is a constant and $\omega_0$ is defined as follows, see~\cite{Pedernales20_GM,Marshman:2021wyk}:
\begin{equation}
    \omega_0 = \bigg(-\frac{\chi_\rho}{\mu_0}\bigg)^{{1}/{2}}\eta_0 \label{eq.omega0_exp}
\end{equation}
From eq.\ref{eq.Lag_j}, we obtain the following equation of motion:
\begin{align}
    m\ddot{x}_j(t)&= -m\omega_0^2x_j(t) - C_j\eta_0\label{eq.EOM_det}
\end{align}
where, 
\begin{equation}
    C_j = \big( S_{xj}\hbar\gamma_e - \frac{ \chi_\rho m}{\mu_0}B_0\big)
\end{equation}
Imposing $x_j(0) = 0$ and $\dot{x} = 0$, we get
\begin{equation}
    x_j(t) = \frac{C_j \eta_0}{m\omega_0^2 } (\cos(\omega_0 t) - 1) \label{eq.det_traj}
\end{equation}
Now, we wish to probe the effect of noise in the magnetic field in this setup. Hence, we consider the magnetic field to have a time-dependent component arising from the noise. Thus consider, along the $x$-direction: 
\begin{equation}
B(t) = B_0 + \left(\eta_0\,+\,\delta\eta_0(t)\right) x.
\end{equation}
where $\delta\eta_0(t)$ represent the noise in the magnetic field gradient~\footnote{We do not include the fluctuations in the magnetic field since we focus on the noise due to the source creating the magnetic field gradient - a wire-like approximation for the current source. However the same procedure can be extended to include the noise in the bias field, say due to a Helmholtz coil current source.}.
The noise is characterized by the ensemble average over time of its various time correlation functions: $E[\delta\eta(t)]$, $E[\delta\eta(t)\delta\eta(t')]$, see~\cite{Milburn:2015}. We shall focus on two types of noise: white noise and flicker noise. 

In a closed-loop interferometric experiment involving a spin superposition, the spin readout is typically performed after the interferometric path is closed to extract information about the evolution of the constituent spin states during the experiment. Our objective is to study how noise influences the relative phase accumulation between these spin components. In addition to phase fluctuations, noise also perturbs the spatial trajectory of the particle, which we denote by \( \delta x_j \). Thus, the effect of noise relevant to us is encoded in the difference in the phases between the spin states arising due to the noise:
\begin{align}
    \delta \phi = \frac{1}{\hbar}\int_0^{T}(&L_R(t) - L_L(t))\,dt\\
    =\frac{1}{\hbar}\int_0^T \Bigg[&-\frac{1}{2}m\bigg(-\frac{\chi_\rho}{\mu_0}\bigg)2\eta_0\delta\eta(t)(x_R^2-x_L^2) \nonumber\\
    &- \frac{1}{2}m\bigg(-\frac{\chi_\rho}{\mu_0}\bigg)\eta_0^2(2x_R\delta x_R-2x_L\delta x_L) \nonumber\\
    &- \hbar\gamma_e\delta\eta(t) (x_R+x_L) - \hbar\gamma_e\eta_0 (\delta x_R+\delta x_L) \nonumber \\
    &+ \frac{ \chi_\rho m}{\mu_0}B_0\delta\eta(t)(x_R-x_L)
    \nonumber \\
    &+ \frac{ \chi_\rho m}{\mu_0}B_0\eta_0(\delta x_R-\delta x_L)\Bigg]\,dt \label{eq.pert_action_complete}
\end{align}
where $L_j(t)$ denote only the time-dependent parts of $L_j$~\footnote{Had we included fluctuations in the bias field, we would have an additional term like $(S_{xR}-S_{xL})\hbar\gamma_e\delta B_0$.}.

We have ignored terms with higher-order dependence on deviations due to noise - higher order in $\delta\eta$ and $\delta x_j$. In the present analysis, we shall ignore the contribution of the fluctuations in the trajectory; the latter is considered in Appendix.\ref{appendix:D}. In effect, we will be considering:
\begin{align}
    \delta \phi =\frac{1}{\hbar}\int_0^T \Bigg[&-\frac{1}{2}m\bigg(-\frac{\chi_\rho}{\mu_0}\bigg)2\eta_0\delta\eta(t)(x_R^2-x_L^2) \nonumber\\
    &- \hbar\gamma_e\delta\eta(t) (x_R+x_L) \nonumber \\
    &+ \frac{ \chi_\rho m}{\mu_0}B_0\delta\eta(t)(x_R-x_L)
    \Bigg]\,dt \label{eq.pert_action}
\end{align}
We now Fourier transform to the frequency domain, substitute the trajectory using eq.\ref{eq.det_traj} and simplify further. 

\begin{align}
    \delta \phi
    =H&\bigg(\int^{-\omega_{min}}_{-\infty}+\int_{\omega_{min}}^{\infty}\bigg)\Bigg[\delta\tilde{\eta}(\omega)\nonumber\\
    &\int_0^T\bigg\{ \frac{(\cos(\omega_0 t) - 1)\cos (\omega_0 t)}{\omega_0^2} \,\bigg\}e^{i\omega t}\,dt\Bigg]\,d\omega \label{eq.pert_action_subed}
\end{align}
where,
\begin{equation}
    H=\frac{1}{\hbar}(4\hbar\gamma_eB_0\eta_0\frac{ \chi_\rho}{\mu_0}) = 4\gamma_eB_0\eta_0\frac{ \chi_\rho}{\mu_0} \label{eq.H_exp}
\end{equation}
\noindent 

In this process of simplification from \ref{eq.pert_action} to eq.\ref{eq.pert_action_subed}, it was noticed that the effect of magnetic field perturbations on the interaction of the spin with the external magnetic field gradient and that of the diamagnetic environment with the external magnetic field gradient are equal, i.e. second and the third terms in eq.\ref{eq.pert_action} are equal upon substituting for $x_j$ with eq.\ref{eq.det_traj}.

In the following analysis, we compute \( \Gamma \), as defined in eq.\ref{eq.Dephasing_genform}, for both white and flicker noise spectra in a simple Stern–Gerlach-type interferometer. Based on the resulting dephasing, we then place constraints on the parameters characterizing the noise source.



\section{Noise in Basic One Loop Interferometer}\label{sec.4}

In this section, we will be computing the values of dephasing under different noise statistics. We wish to see what should be the maximum values of $A^2$ in case of white noise - eq.\ref{eq.whitenoise_statprop2}, and K in case of flicker noise - eq.\ref{eq.PSD_flicker} so that the coherence at the end of the interferometer, due to these noises is around $10\%$. We also study the trend of the dephasing rate under various experimental parameters and hence we can extrapolate to other values of coherence. Since we want an order of magnitude estimate, we carry out a simple first analysis for a single loop interferometer and do not consider time-dependent spin states (as would be the case in the QGEM protocol). That is we just consider a simple harmonic oscillator case given by eq.\ref{eq.Lag_j}

Using eq.\ref{eq.pert_action_subed}, we obtain the dephasing for a general noise PSD to be:

\begin{align}
    \Gamma &= \frac{8H^2}{\omega_0^5} \int_{{\omega_{\text{min}}}/{\omega_{0}}=1}^{\infty} S_{\eta\eta}(\xi)F_{HO}(\xi) d\xi \label{eq.G_d}
\end{align}
where we consider general noise statistics according to eq.(\ref{eq.gen_stat_noise}), and $\xi = {\omega}/{\omega_0}$, and $F_{HO}(x)$ is given by:

\begin{align}
    F_{HO}(\xi)&= \sin^{2}\left(\pi \xi\right) \left[ \dfrac{1}{2\xi} + \dfrac{\xi}{2\left(\xi^{2} - 4\right)} - \dfrac{\xi}{\xi^{2} - 1} \right]^{2}\nonumber\\
    &= \frac{1-\cos(2\pi \xi)}{2} \left[\frac{\xi^2+2}{\xi(\xi^2-4)(\xi^2-1)}\right]^{2}\label{eq.FHO_transferfnt}
\end{align}
We can consider eq.\ref{eq.FHO_transferfnt} to represent the effective transfer function~\footnote{It would be more meaningful to consider ${4H^2}/{\omega_0^5}F_{HO}$ to be the effective transfer function, as it captures the system dependent parameters via $H$ and $\omega_0$.} for a harmonic oscillator up to a system constant (and under the assumption of ignoring deviations in the trajectory). Note, that the transfer function is an even function in $\xi$, and hence the integral over negative frequencies is the same as that of positive values. Next, we shall compute the integral over $\omega$ using numerical methods, for various forms of $S(\omega)$.

We first compute the parameter values expected in a typical experiment, so that we can find the bounds on the order of magnitude of noise. These parameters are motivated by the experimental setup of levitating nanoparticles in a diamagnetic trap, see~\cite{Elahi:2024dbb}.

\subsection{Estimation of parameters}\label{sec.a}

The values of parameters considered in our analysis are presented in Table \ref{tab:parameters}. The values of the current and the distance from the wire are motivated from  \cite{Elahi:2024dbb}.
\begin{table}[h]
    \centering
    \begin{tabular}{|l|c|}
        \hline
        \textbf{Parameter} & \textbf{Value} \\
        \hline
        Electron gyromagnetic ratio, $\gamma_e$ & $1.761 \times 10^{11} \, \text{s}^{-1} \, \text{T}^{-1}$ \\
        Bias magnetic field, $B_0$ & $0.2$ T \\
        Current, $I$ & $12$ A \\
        Distance from wire, $d$ & $20 \, \mu$m \\
        Nanodiamond density, $\rho$ & $3.5 \times 10^3 \, \text{kg} \, \text{m}^{-3}$ \\
        Mass Magnetic susceptibility, $\chi_\rho$ & $-6.286 \times 10^{-9} \, \text{m}^{3} \, \text{kg}^{-1}$ \\
        \hline
    \end{tabular}
    \caption{List of parameters used in the calculation.}
    \label{tab:parameters}
\end{table}
Next, we derive the key quantities $\eta_0$, $H$ (see, eq.~(\ref{eq.H_exp})), $\omega_0$, and the maximum spin state separation $\Delta x_{\text{max}}$, using the parameter values from Table \ref{tab:parameters}. We can evaluate the magnetic field gradient using eq.\ref{eq.eta0_exp}:  
    \begin{equation}
        \eta_0 = -\frac{\mu_0 I}{2\pi d^2} = -0.6\times10^{4} \,\text{T}\,\text{m}^{-1}.\label{eq.eta0}
    \end{equation}
    We can compute $H$ using eq.\ref{eq.H_exp}: 
    \begin{equation}
        H = 4 \gamma_e B_0 \eta_0 \frac{\chi_\rho}{\mu_0} = 4.23\times10^{12} \,s^{-1} \text{m}\, \text{kg}^{-1} \text{A}^{-2} \label{eq.H}
    \end{equation}
and $\omega_0$, using eq.\ref{eq.omega0_exp}:
    \begin{equation}
        \omega_0 = \bigg|\left(-\frac{\chi_\rho}{\mu_0}\right)^{\frac{1}{2}} \eta_0\bigg| = 4.24\times10^{2} \text{ Hz}.\label{eq.omega0}
    \end{equation}
Thus the experimental time, that is the time taken to complete the one-loop interferometer is given by:
\begin{equation}
    T_{\text{exp}}=\frac{2\pi}{\omega_0} = 1.48\times10^{-2}\,\,\text{s}\,.
\end{equation}
Hence, the Coherence is measured by $e^{-\Gamma T_{exp}}$:
Thus, for a coherence of 10\%, we get, 
\begin{equation}
    \Gamma \approx 155 \,\text{Hz}^{-1} \,.
\end{equation}
The maximum separation of the spin states occurs at half the time period of the oscillation($t={\pi}/{\omega_0}$) and is found from eq.\ref{eq.det_traj}. The final simplified form of the maximum separation of spin states is given by:
    \begin{align}
    \Delta x|_{Max} &= \bigg|x_R(t)-x_L(t)\bigg|_{t =\frac{\pi}{\omega_0}}  \nonumber\\
    &=\bigg|\frac{4\hbar\gamma_e\eta_0}{m\omega_0^2}\bigg|  \approx 2.5 \times10^{-9}m\,.\label{eq.max_sep}
\end{align}


\subsection{White Noise}

We carry out the $\omega$ integrals in eq.\ref{eq.G_d} using numerical integration with the PSD for white noise (eq.\ref{eq.whitenoise_statprop2}):
\begin{equation}
    S_{\eta\eta}(\xi) = A^2
\end{equation}
The numerical analysis reveals that, despite the frequency-independent nature of the power spectral density (PSD), the effective transfer function (Eq.~\ref{eq.FHO_transferfnt}) ensures that the resulting dephasing remains finite for a given value of \( A \). A plot of the transfer function is provided in Appendix~\ref{appendix:E}. This behaviour indicates that the harmonic oscillator is more sensitive to low-frequency noise components than to high-frequency ones. 

The following calculation is helpful for any harmonic oscillator potential. We take $T_{exp} = {2\pi}/{\omega_0}$ since we are considering a single loop interferometer ~\footnote{However, to obtain the phase information, we have to carry out about $10^4$ to $10^6$ experimental runs. Hence, the precision of the measurement increases with the number of experimental runs. However, this also increases the sensitivity to noise. Hence we need to consider a $T_{exp} = 10^4 \times \frac{2\pi}{\omega_0}$ if we consider $10^4$ experimental runs. This increases the bandwidth of frequencies to which the experiment is sensitive since $\omega_{min} = \frac{2\pi}{T_{exp}}$. \label{footnote.exptime}}. Now we impose the bounds on $\omega$: the frequency of the noise has to be greater than the inverse of the total time of the experiment (as discussed before). Thus, we take $\omega_{min} = \omega_0$. From eq.\ref{eq.G_d}, we carry out the integration numerically and obtain ~\footnote{Since the integrand in eq.(\ref{eq.WN_simple}) is well defined for all frequencies (refer to Appendix.\ref{appendix:E}), we can obtain the upper bound on dephasing by considering $\omega_{min} = 0$. This gives us: $\int_{0}^{\infty} F_{HO}(\xi) d\xi \approx 4.3$ and the upper bound on dephasing is $\Gamma_W \approx \frac{8H^2}{\omega_0^5} A^2 \times4.3$. \label{footnote.integrand_WN}}:
\begin{align}
    \Gamma_{W} &= \frac{8H^2}{\omega_0^5} A^2\int_{1}^{\infty} F_{HO}(\xi) d\xi \nonumber\\
    \implies\Gamma_W &\approx \frac{8H^2}{\omega_0^5} A^2 \times1.8 \,\text{s}^{-1}\label{eq.WN_simple}
\end{align}
where $H = 4 \gamma_e B_0 \eta_0 \frac{\chi_\rho}{\mu_0}$, as defined in eq.\ref{eq.H_exp}.

\begin{figure}
    \centering
    \includegraphics[width=\linewidth]{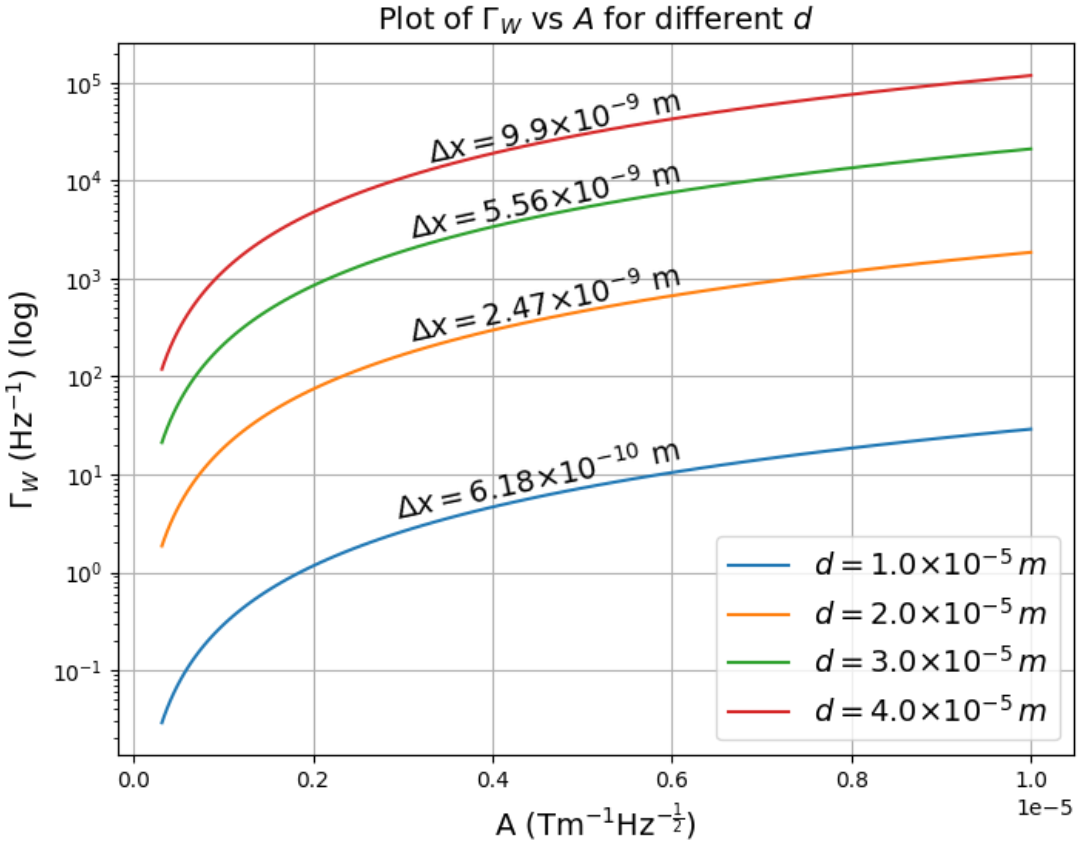}
    \captionsetup{justification=RaggedRight}
    \caption{Dependence of the dephasing rate \( \Gamma_W \), induced by white noise in the current generating the magnetic field gradient, on the noise amplitude $A$, for various distances $d$ between the NV-diamond and the wire. The current and the values of $d$ are motivated from a levitating nanodiamond diamagnetically, via chip configuration, see \cite{Elahi:2024dbb}.
    The remaining parameters follow the simplified Stern-Gerlach model for creating spatial superposition (see Section\ref{sec.a}). The quantity \( \Delta x \), defined in Eq.(\ref{eq.max_sep}), represents the maximum spatial separation between the spin components of the superposition state. The plot illustrates how both the magnitude of current fluctuations and the spatial configuration significantly affect spin coherence in a Stern–Gerlach–type interferometer. We have ensured here that $\Gamma_{W} T_{\rm exp}\leq 2.5$ for d=20 $\mu$m.  } 
    \label{fig:WN for dif d gvA}
\end{figure} 
We observe from Fig.~(\ref{fig:WN for dif d gvA})that increasing the distance \( d \) between the NV-diamond and the wire leads to a decrease in dephasing \( \Gamma_W \) for a given noise amplitude \( A \). Physically, this can be attributed to the reduction in the magnetic field gradient with increasing distance from the wire. Interestingly, a stronger magnetic field gradient leads to a smaller spatial separation \( \Delta x \) between the spin states. This counterintuitive behaviour arises because the effective harmonic oscillator frequency \( \omega_0 \) increases with the field gradient, thereby increasing the confining potential and reducing the spatial spread of the wavepacket.

This introduces a design trade-off: while a smaller magnetic field gradient increases susceptibility to current noise, it also results in a larger superposition size, which enhances the interferometric signal. Also, other noise sources like the electromagnetic field noise \cite{Afek:2021bua, Fragolino:2023agd,Munday:2020} contribute nearer to the current carrying wire. Hence, we have to choose an optimal magnetic field gradient such that the noise is low but the superposition size gives us measurable effects. 

From eq.(\ref{eq.WN_simple}), we find that for a tolerable dephasing rate of \( \Gamma_W = \Gamma_{W_{\text{max}}}\), $A$ can be constrained using:
\begin{equation}
    A\lesssim \sqrt{ \Gamma_{W_{\text{max}}} \times \frac{\omega_0^5}{8H^2\times1.8}} \label{eq.WN_bound_exp}
\end{equation}
Specifically, if we demand that the dephasing rate should be less than $\sim100$ Hz (i.e, \(\Gamma_{W_{\text{max}}} \sim 100\)Hz) and using the values of \( H \) and \( \omega_0 \) from eqs.~(\ref{eq.H} and \ref{eq.omega0}), respectively, the noise amplitude is constrained as:
\begin{align}
    A &\lesssim 2.9 \times 10^{-6} \,\, \text{T}\,\text{m}^{-1}\,\text{Hz}^{-1/2} \label{eq.WN_Bound}
\end{align}
This allows us to estimate the maximum permissible ratio of current noise~\footnote{\label{footnote.WFN}The noise magnitude is given by \( \sqrt{\int_{\omega_0}^{\infty} S(\omega) F_{\text{HO}}(\omega)\,d\omega} \). This can be intuitively understood as the square root of power, where power is the product of the power spectral density (PSD) and the effective bandwidth.} to signal current for a dephasing rate of $\sim{10^2}$ Hz using eq.(\ref{eq.eta0_exp}):
\begin{align}
    \frac{\delta I}{I} \approx \frac{A\sqrt{1.8\,\omega_0}}{\eta_0} = 1.3 \times 10^{-8}
\end{align}
In the above we have taken the configuration where  $|\eta_0|\sim 6\times 10^{3} \,{\text{Tm}^{-1}}$. 

If we consider noise sensitivity for the total experimental time (say $10^4\times\frac{2\pi}{\omega_0}$), then the bound on $A$ must be re-evaluated accordingly. Under this consideration, the frequency bandwidth to which the experiment is sensitive to increases\footref{footnote.exptime}, for example corresponding to $T_{exp} = 10^4\times\frac{2\pi}{\omega_0}$, the $\omega_{min} = 10^{-4}\times\omega_0$. This further varies the bounds of the integral\footref{footnote.integrand_WN} in eq.(\ref{eq.WN_simple}). Hence eq.\ref{eq.WN_bound_exp} would change accordingly. The bound on $A$ becomes stricter, such as for $T_{exp} = 10^4\times\frac{2\pi}{\omega_0}$, $A\lesssim0.7\times10^{-6} \, \text{T}\,\text{m}^{-1}\,\text{Hz}^{-1/2}$. The increase in bandwidth and stricter condition on $A$ compensate for each other giving back the same tolerance for current fluctuations. This is further explained by eq.(\ref{eq.dI/I}).

As discussed earlier in this section, the convergence of \( \Gamma_W \) for flat (white) noise spectra arises due to the frequency-dependent suppression built into the harmonic oscillator's transfer function. This indicates that the harmonic oscillator is particularly sensitive to low-frequency noise. Since current noise in conductors typically exhibits a \( 1/f \) (flicker) behaviour, we now turn to an analysis of flicker noise in the next section.

\subsection{Flicker Noise}
In the case of Flicker noise, the PSD in terms of the dimensionless quantity $\xi$ is given by \cite{PhysRevB.32.736, RevModPhys.53.497}:
\begin{equation}
    S_{\eta\eta}(\xi) =\frac{\mu_0}{2\pi d^2} \frac{KI^2}{\abs{\omega_0}^\alpha|\xi|^\alpha} \equiv \frac{\tilde K^2(\omega_0)}{|\xi|^\alpha}\label{eq.PSD_flicker_x}
\end{equation}
where $\tilde K$ is $\tilde K = \sqrt{\frac{\mu_0}{2\pi d^2} \frac{KI^2}{\abs{\omega_0}^\alpha} }$. We once again take $T_{exp} = {2\pi}/{\omega_0}$ since we are considering a single loop interferometer and we numerically carry out the $\omega$ integral.
From eq.(\ref{eq.G_d}), we obtain~\footnote{The integrand in eq.(\ref{eq.FN_simple}) is not well defined at $\xi =0$ (refer to Appendix.\ref{appendix:E}), hence to obtain the upper bound on dephasing we need to consider a large but finite number of experimental runs. If we consider $10^4$ experimental runs, the total experimental time is given by $T_{exp} = 10^4\times \frac{2\pi}{\omega_0}$. Thus $\frac{\omega_{min}}{\omega_0}= 10^{-4}$. This gives us: $\int_{10^{-4}}^{\infty} F_{HO}(\xi) d\xi \approx 24$ and the corresponding dephasing is $\Gamma_F \approx \frac{8H^2}{\omega_0^5} \tilde{K}^2 \times24$. Similarly, if we consider $10^6$ experimental runs, the dephasing is $\Gamma_F \approx \frac{8H^2}{\omega_0^5} \tilde{K}^2 \times 35.5$. Therefore the increase in the magnitude of dephasing is utmost of order $10^1$.\label{footnote.integrand_FN}}:
\begin{align}
    \Gamma_{F} &= \frac{8H^2}{\omega_0^5} \tilde K^2(\omega_0)\int_{1}^{\infty} \frac{F_{HO}(\xi)}{|\xi|} d\xi \nonumber\\
    \implies\Gamma_F &\approx \frac{8H^2}{\omega_0^5} \tilde K^2(\omega_0) \times1.3  \,\,\, \text{s}^{-1}\label{eq.FN_simple}
\end{align}

\begin{figure}
    \centering
    \includegraphics[width=\linewidth]{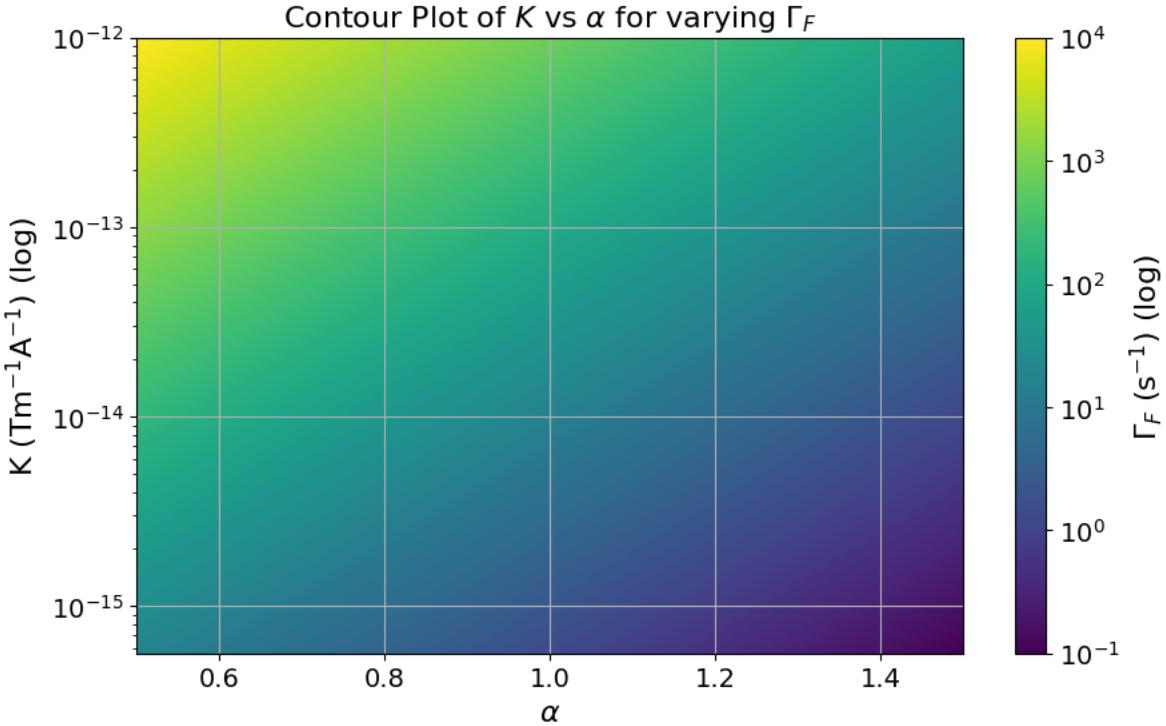}
    \captionsetup{justification=RaggedRight}
    \caption{Contour plot showing the behaviour of the source dependent component of the flicker noise, \( K \) as a function of the dimensionless parameter \( \alpha\in[0.5,1.5] \) that determines the dependence of noise statistics on frequency, with the contours representing varying levels of the resulting dephasing rate\( \Gamma_F\). \( \omega_0 \), is the characteristic frequency of the system, defined in eq.~(\ref{eq.omega0}). Smaller \( \alpha \) causes stronger dephasing for the same value of \( K \). This shows that the system is particularly sensitive to low-frequency noise, which is a characteristic trait of flicker noise. The plot helps visualize how different combinations of $K$ and $\alpha$ contribute to decoherence in the system, and it can be used to identify safe operating regions where dephasing remains within acceptable bounds.}
   \label{fig:flickernoise_HOom5}
\end{figure}

\begin{figure}
    \centering
    \includegraphics[width=\linewidth]{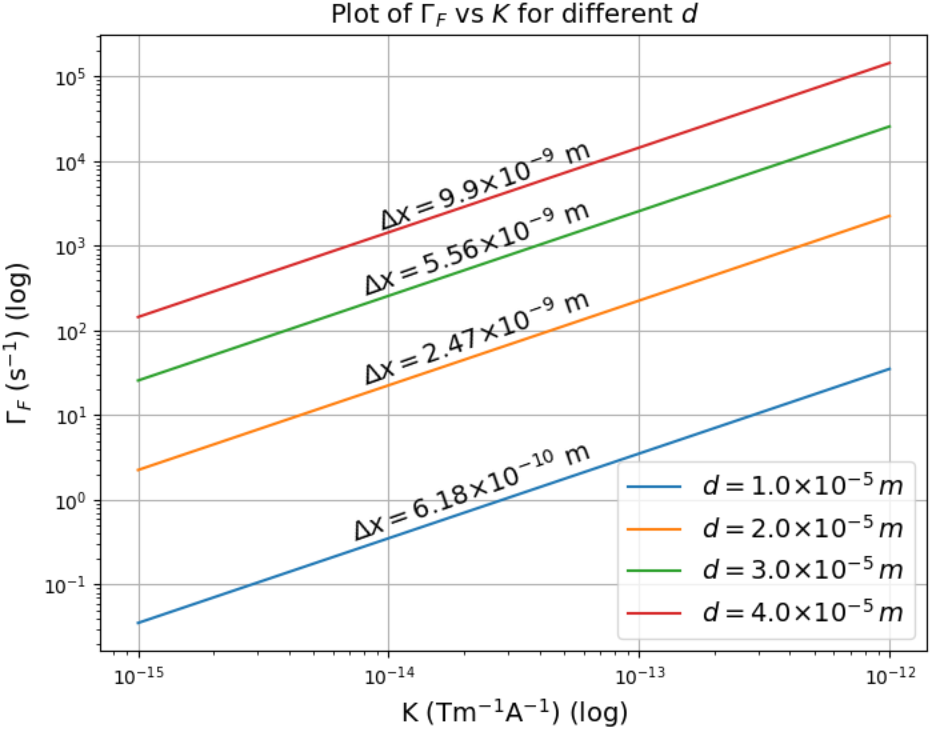}
    \captionsetup{justification=RaggedRight}
    \caption{Dependence of the dephasing rate \( \Gamma_F \), induced by flicker (1/f) noise in the current generating the magnetic field gradient, on the source-dependent noise amplitude \( K \), for various distances \( d \) between the NV-diamond and the wire. We have fixed $\alpha = 1$ since this is the approximate value of the exponent in the case of superconductors. The parameter \( K \) depends on the material properties of the superconducting wire, such as temperature, composition, and fabrication details. All other parameters are fixed as specified in Sec.~\ref{sec.a}, consistent with the simplified Stern-Gerlach model. The quantity \( \Delta x \) denotes the maximum spatial separation between the spin components of the superposition state, as defined in Eq.~(\ref{eq.max_sep}). This figure illustrates the sensitivity of spin coherence to both current noise magnitude and the geometric configuration of the Stern-Gerlach type interferometer.}
   \label{fig:flickernoise}
\end{figure}

From Fig.\ref{fig:flickernoise_HOom5}, we observe that the greater the values of $K$, the higher the dephasing, as expected. As the spectral exponent $\alpha$ increases, the contribution of high-frequency components in the noise spectrum diminishes, making the noise increasingly dominated by slow, quasi-static fluctuations. These low-frequency drifts behave effectively as static offsets in system parameters over the experimental timescale (as discussed in Sec.\ref{sec.fluctuations and dephasing}). Consequently, their impact on the system's coherence is reduced, since such slowly varying fluctuations can often be compensated for or averaged out. For this reason, their contribution to dephasing is neglected in this analysis.
For superconductors, $\alpha\approx1$, and for this value of the spectral exponent, we observe the trend of the dephasing rate as a function of source-dependent $K$. From Fig.(\ref{fig:flickernoise}), we observe that dephasing increases with increasing distance \( d \) from the wire. Specifically, doubling \( d \) results in an approximate two-order-of-magnitude increase in the dephasing rate \( \Gamma_F \). This trend arises because the magnetic field gradient generated by the current decreases with distance from the wire. A weaker magnetic field gradient leads to a reduced harmonic confinement frequency. Since the amplitude of flicker noise is inversely proportional to the harmonic frequency, a lower frequency results in increased noise.  Physically, a reduced harmonic frequency corresponds to weaker confinement, allowing a larger spatial separation \( \Delta x \) between the spin components of the superposition state. This enhanced separation increases the system’s sensitivity to magnetic field gradient fluctuations.
%
%

%
From eq.(\ref{eq.FN_simple}), we find that for a tolerable dephasing rate of \( \Gamma_F = \Gamma_{F_{\text{max}}}\), $\tilde{K}$ can be constrained using:
\begin{equation}
    \tilde{K}\lesssim \sqrt{ \Gamma_{F_{\text{max}}} \times \frac{\omega_0^5}{8H^2\times1.3}} \label{eq.FN_bound_exp}
\end{equation}
Specifically, if we demand that the dephasing rate should be less than $\sim100$ Hz (i.e, \(\Gamma_{F_{\text{max}}} \sim 100\)Hz) and using the values of \( H \) and \( \omega_0 \) from eqs.~(\ref{eq.H} and \ref{eq.omega0}), respectively, and further using relation between $K$ and $\tilde{K}$ from eq.\ref{eq.PSD_flicker_x}, we can constrain $K$: 
\begin{align}
    K&\lesssim 0.7\times10^{-13} \,\text{T\,m$^{-1}$A$^{-1}$}\label{eq.FN_Bound}
\end{align}
$K$ is a source-dependent constant, we note that values of similar magnitude have been reported for niobium (Nb) superconductors in previous studies \cite{WellstoodUrbinaClarke-2004}~\footnote{Specifically,\cite{WellstoodUrbinaClarke-2004} defines a related source-dependent parameter $C$, which can be related to $K$ in our paper as: $C = {K\mathcal{A}}/{T^2}$, where $\mathcal{A}$ is the cross-sectional area of the superconducting wire and $T$ is the temperature. In our source, $\mathcal{A} \approx 7.85\times10^{-11}\text{m}^2$ and $T\approx 4.2\text{K}$. Thus, we estimate $C\approx 0.3\times10^{-23} \,\text{m}^2 \text{K}^{-2}$ in our case. This value is one order of magnitude smaller than
reported value $C=3.9\pm0.4 \times10^{-23}\,\text{m}^2 \text{K}^{-2}$ in 2004.  However, given advances in fabrication and noise mitigation, the current state-of-the-art may exhibit lower values of $C$, and our estimate could be within achievable limits.}.

From this, we can estimate the tolerable ratio of the noise~\footref{footnote.WFN} and signal amplitudes for the magnetic field gradient and also obtain it in terms of current amplitudes from eq.(\ref{eq.eta0_exp}) at $\omega_0=424$~Hz: 
\begin{align}
    \frac{\delta I}{I} \approx \frac{\tilde K\times\sqrt{1.3\,\omega_0}}{\eta_0} = 1.3\times10^{-8}
\end{align}
where $\tilde K$ is defined in eq.(\ref{eq.PSD_flicker_x}).

If we consider noise sensitivity for the total experimental time (say $10^4\times\frac{2\pi}{\omega_0}$), then the bound on $K$ must be re-evaluated accordingly. Under this consideration, the frequency bandwidth to which the experiment is sensitive to increases\footref{footnote.exptime}, for example corresponding to $T_{exp} = 10^4\times\frac{2\pi}{\omega_0}$, the $\omega_{min} = 10^{-4}\times\omega_0$. This further varies the bounds of the integral\footref{footnote.integrand_FN} in eq.(\ref{eq.FN_simple}). Hence eq.\ref{eq.FN_bound_exp} would change accordingly. The bound on $K$ becomes stricter, such as for $T_{exp} = 10^4\times\frac{2\pi}{\omega_0}$, $K\lesssim0.04\times10^{-13} \, \text{T}\,\text{m}^{-1}\,\text{A}^{-1}$. The increase in bandwidth and stricter condition on $K$ compensate for each other giving back the same tolerance for current fluctuations. This is further explained by eq.(\ref{eq.dI/I}).


\subsection{Comparison}
\begin{figure}
    \centering
    \includegraphics[width=\linewidth]{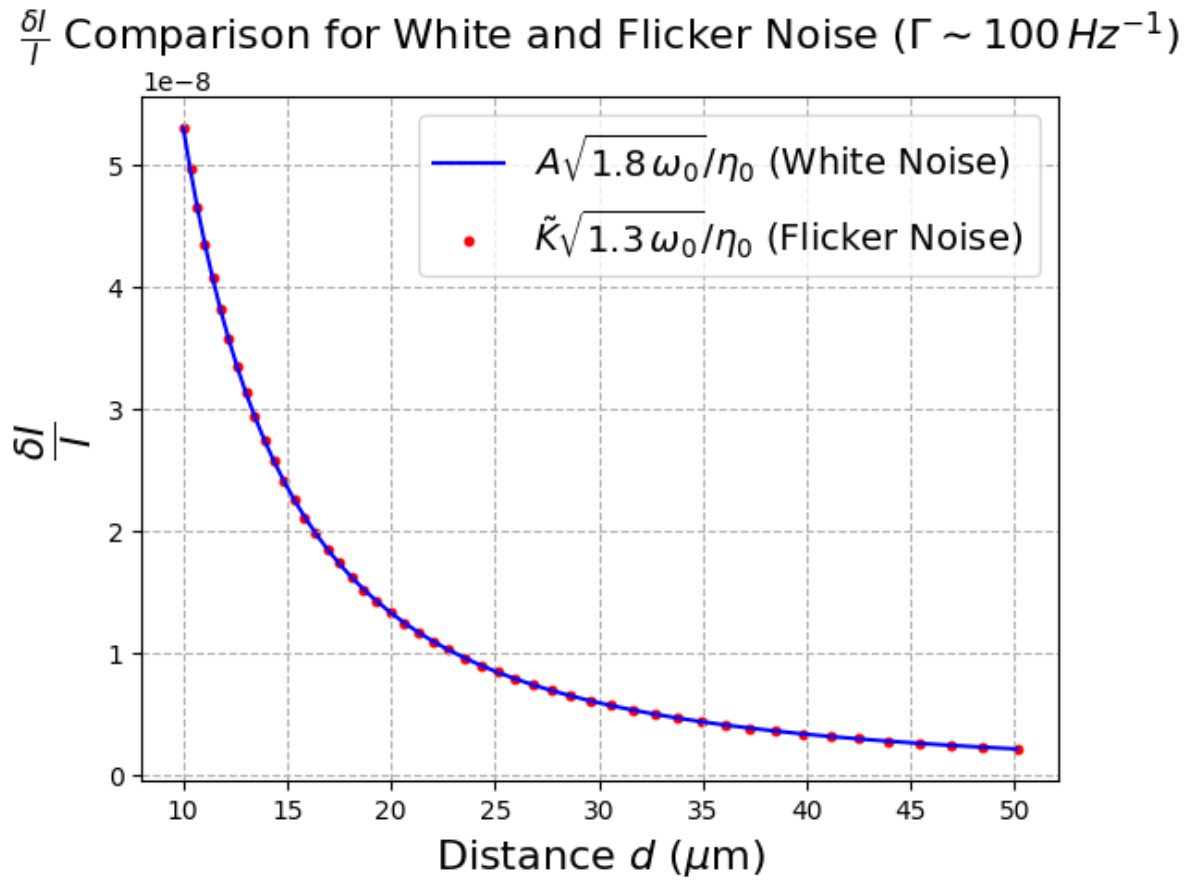}
    \captionsetup{justification=RaggedRight}
    \caption{Comparison of the upper bound on the relative current noise \( \frac{\delta I}{I} \) as a function of the particle’s distance \( d \) from a current-carrying superconductor, for a dephasing rate of \( \Gamma \sim 10^2 \text{ Hz} \). The plot includes results for white noise (blue curve) and flicker noise (red dots). As the distance of the particle from the wire increases, the tolerance for noise drops sharply. Despite differing spectra, both noise types impose similar constraints on current fluctuations over the entire distance range, indicating comparable impacts on coherence in this regime.}
   \label{fig:SNR_comp}
\end{figure}

In Fig.(\ref{fig:SNR_comp}), we compare the upper bound on the noise-to-signal ratio of the current source \( (\delta I / I) \) required to maintain a dephasing rate of \( \sim 10^{2} \) Hz for the two types of noise considered in this study: flicker noise and white noise. We find that the bounds on \( \delta I / I \) are nearly identical in both cases. While the dephasing rate associated with white noise is inherently larger than that of flicker noise, both noise types exhibit comparable tolerance to current fluctuations.

This observation suggests a potentially broader implication: for any spectral exponent \( \alpha \in [0, 1.5] \), the constraints imposed on experimental parameters by dephasing considerations remain approximately constant. Therefore, the requirement on the magnitude of current fluctuations to achieve a given dephasing rate $\sim 10^2$ Hz appears to be relatively robust across different noise spectra. This is because current and magnetic field gradient fluctuations are directly related by eq.(\ref{eq.eta0_exp}) and hence, putting a bound on the dephasing rate obtained from magnetic field gradient fluctuations, directly yields the bounds on both magnetic field gradient and current fluctuations from eq.(\ref{eq.WN_simple}) and eq.(\ref{eq.FN_simple}). Hence, by definition of noise amplitude \footref{footnote.WFN} and from eq.(\ref{eq.G_d}),
\begin{equation}
    \frac{\delta I}{I} = \frac{\delta \eta}{\eta_0} = \sqrt{\frac{\Gamma}{2}} \frac{\omega^3_0}{2 H\eta_0} \label{eq.dI/I}
\end{equation}
Thus, given an upper bound on the dephasing, the tolerance of current fluctuations is determined by system parameters and not by noise parameters nor the experimental time; i.e., even if we take the total experimental time including all experimental runs ($\approx10^4\times\frac{2\pi}{\omega_0}$), it imposes constraints only on the noise parameters like $A$ in case of white noise and $\tilde{K}$ in case of flicker noise, but not on the bound of $\frac{\delta I}{I}$.


We also observe that as the distance between the particle and the wire increases, the constraint on the noise-to-signal ratio becomes more stringent in order to maintain low dephasing. This is due to the reduction in the magnetic field gradient $\eta$ at larger distances, which leads to weaker confinement of the particle, see \cite{Elahi:2024dbb}. As a result, the system becomes more susceptible to noise-induced fluctuations, thereby requiring a lower noise-to-signal ratio to preserve coherence.

\section{Humpty-Dumpty Problem}\label{sec.5}

In Sec.\ref{sec.fluctuations and dephasing} we discussed the general form of noise analysis in cases where the deviations in the trajectory due to the noise are ignored. Now we shall analyze the impact of the deviations in the trajectory (indicated by $\delta x_j$ in eq.(\ref{eq.pert_action_complete}) due to fluctuations in the magnetic field gradient and quantify this effect with the contrast of the interference. Contrast, $\mathcal{C}$ in an interferometer experiment lies in the range [0,1], where $\mathcal{C} = 1$ means that the interferometer closes and $\mathcal{C}=0$ means that the interferometer doesn't close and hence spin-readout is not possible, see the definition below. The latter is part of the Humpty-Dumpty problem of irreversibility associated with quantum measurement, specifically the practical impossibility of restoring a coherent superposition once decoherence has occurred. We analyse the value of $\mathcal{C}$ subject to the parameters obtained above, see~\cite{Englert,Schwinger,Scully,Margalit:2020qcy}. In the Appendix.\ref{appendix:E}, we derive the dephasing when the deviations in the trajectory are accounted for in eq.(\ref{eq.pert_action_complete}).

\subsection{Equations of Motion}

We start with analysing the equation of motion of the deviations from the trajectory $\delta x_j$. They satisfy the following equation (obtained from eq.(\ref{eq.def_dev_trajd}, \ref{eq.Lag_j} and \ref{eq.EOM_det})):
\begin{align}
    \delta \ddot x_j(t) + \omega_0^2\delta x_j(t)  = -\frac{C_j \delta\eta(t)}{m} (2\cos(\omega_0 t) - 1)\label{eq.devtraj_EOM}
\end{align}
To solve this, we obtain:
\begin{equation}
\delta x_{j}(t) = -\frac{C_j}{m\omega_0} \int_{0}^{t} dt' \, \left( 2 \cos(\omega_0 t') - 1 \right)\sin\left[ \omega_0 (t - t') \right] \, \delta \eta(t')  \label{eq.dev_traj_t}
\end{equation}
We can also obtain the solutions in the Fourier domain which would help evaluate eq.(\ref{eq.G_d}).
\begin{equation}
    {\delta \tilde x_j}(\omega )\ =\frac{C_j}{m(\omega^2-\omega_0^2)}\bigg(\delta\tilde{\eta}(\omega-\omega_0) + \delta\tilde{\eta}(\omega+\omega_0) - {\delta\tilde{\eta}}(\omega )\bigg)\label{eq.dev_traj_fouriersoln}
\end{equation}
To understand if the one-loop interferometer can be approximated as a closed loop, we also need to understand the evolution of the momentum. 
\begin{align}
    p_j = \frac{\partial{L_j}}{\partial{\dot{x}}} = m\dot x_j 
    \implies\delta p_j(t) = m \delta \dot{x}_j(t) \nonumber
\end{align}

Thus, from eq.(\ref{eq.EOM_det}) , we obtain:
\begin{equation}
    p_j(t) = -\frac{C_j \eta_0}{\omega_0 } \sin(\omega_0 t) \label{eq.det_mom}
\end{equation}
From eq.(\ref{eq.devtraj_EOM}), we obtain:
\begin{equation}
    \implies\delta p_j(t)= -C_j \int_{0}^{t} dt' \bigl(2\cos(\omega_0 t') - 1\bigr)\cos\bigl[\omega_0 (t - t')\bigr] \, \delta \eta(t') \, 
\end{equation}
where, $p_j^{tot}(t) = p_j(t) + \delta p_j(t)$. In the Fourier domain:
\begin{equation}
    {\delta \tilde p_j}(\omega )\ =-\frac{i \omega C_j }{m(\omega^2-\omega_0^2)}\bigg(\delta\tilde{\eta}(\omega-\omega_0) + \delta\tilde{\eta}(\omega+\omega_0) - {\delta\tilde{\eta}}(\omega )\bigg)\label{eq.dev_mom_fouriersoln}
\end{equation}

\subsection{Calculating the Contrast}
Since we are considering a harmonic oscillator case, we will compute the expected contrast loss for the ground state wavefunction, a Gaussian:
\begin{equation}
    \psi =  \left( \frac{1}{2\pi\sigma^2_0} \right)^{1/4} \text{exp}\left({-\frac{x^2}{4\sigma^2_0}}\right)
\end{equation}
where
\begin{equation}
    \sigma_0 \equiv \sigma_x(t=0) = \sqrt{\frac{\hbar}{2m\omega_0}}
\end{equation}
We consider $\hbar=1$. Contrast is given by~\cite{Englert,Schwinger,Scully,Margalit:2020qcy}:
\begin{equation}
    \mathcal{C}(t) = \exp\bigg(-\frac{1}{2}\bigg[\bigg(\frac{\Delta x(t)}{\sigma_x}\bigg)^2+\bigg(\frac{\Delta p(t)}{\sigma_p}\bigg)^2\bigg]\bigg) \label{eq.Contrast}
\end{equation}
where $\Delta x(t)$ and $\Delta p(t)$ are the differences in position and momentum between the two arms of the interferometer, at time $t$.
\begin{align}
    \Delta x(t) &= x_R^{tot}(t) - x_L^{tot}(t) = \delta x_R(t) - \delta x_L(t)\\
    \Delta p(t) &= p_R^{tot}(t) - p_L^{tot}(t) \nonumber\\
    &= \delta p_R(t) - \delta p_L(t) - \frac{2\hbar\gamma_e \eta_0}{\omega_0} \sin(\omega_0 t)
\end{align}
Note that $\Delta p\sim\order{10^{-12}\delta x}$ due to the small mass of the particle $m\sim\order{10^{-15}}$ kg. Hence the dominant term contributing to $\Delta p$ comes from solving the unperturbed equation of motion $\Delta p\equiv -\frac{2\hbar\gamma_e \eta_0}{\omega_0} \sin(\omega_0 t)$. Therefore, assuming minimum uncertainty, we obtain:
\begin{equation}
    \sigma_p = \frac{1}{2\sigma_x}
\end{equation}
In each run of the experiment, at the final evolution time $T_{exp} = {2\pi}/{\omega_0}$, the interferometer must be closed such that the wavepackets corresponding to different spin components overlap sufficiently in space. However, to extract the full interference pattern via spin readout, data must be accumulated over multiple experimental runs. Consequently, the theoretical estimation of contrast requires analysis at both the single-run and ensemble-averaged levels, as defined in Eq.~(\ref{eq.Contrast}). The single-run contrast depends on a specific realization of the noise, which is generally unknown and must therefore be evaluated through numerical simulation. In contrast, the ensemble-averaged contrast can be computed analytically, as it relies on the known statistical properties of the noise. The ensemble average is given by~\cite{Englert,Schwinger,Scully,Margalit:2020qcy}:
\begin{equation}
    \mathcal{C}(t) = \exp\bigg[-\frac{1}{2}\bigg(\frac{\langle \Delta x(t)^2\rangle}{\sigma_x^2}+4\sigma_x^2\langle \Delta p(t)^2\rangle\bigg)\bigg] \label{eq.Contrast_avg}
\end{equation}
Now we will compute the theoretical values of contrast expected in case of different noise statistics: white noise and flicker noise.

Since we are interested in a one-loop interferometer, thus we take measurements at $ t = T_{exp} = {2\pi}/{\omega_0}$. The width of a Gaussian wavefunction after a one-loop interferometer ideally returns to the original width of the wavefunction, see \cite{Zhou:2024voj}. Hence,
\begin{equation}
    \sigma_x = \sqrt{\frac{\hbar}{2m\omega_0}}
\end{equation}
To evaluate the deviation in trajectory arising from noise in a single experimental run, we numerically simulate the two types of noise, white noise and flicker (1/f) noise, independently. The algorithm generates noise in the frequency domain by first defining the appropriate power spectral density (PSD) corresponding to each noise type. The spectral characteristics are introduced by weighting a complex Gaussian random generator with the square root of the PSD, ensuring that the generated frequency-domain signal exhibits the desired statistical properties. This noise is then used to compute the deviation in each of the trajectories (left and right arms of the interferometer) by solving the perturbed equations of motion in the Fourier domain, eq.(\ref{eq.dev_traj_fouriersoln}) and eq.(\ref{eq.dev_mom_fouriersoln}). The resulting deviations in position and momentum are transformed to the time domain via inverse Fourier transform and are subsequently used to calculate the expected contrast. This enables an assessment of whether the noise-induced deviations in the trajectory are small enough to allow measurement at the end of the interferometer. The values of the parameters used in the algorithm are from Sec.\ref{sec.a} and based on the upper bounds on the source-dependent parameters obtained in the previous sections (eq.(\ref{eq.WN_Bound}) for white noise and eq.(\ref{eq.FN_Bound}) for flicker noise).

\subsubsection*{White noise and Flicker Noise}



\begin{figure}[!ht]
    \centering
    \begin{subfigure}[b]{\linewidth}
        \includegraphics[width=\linewidth]{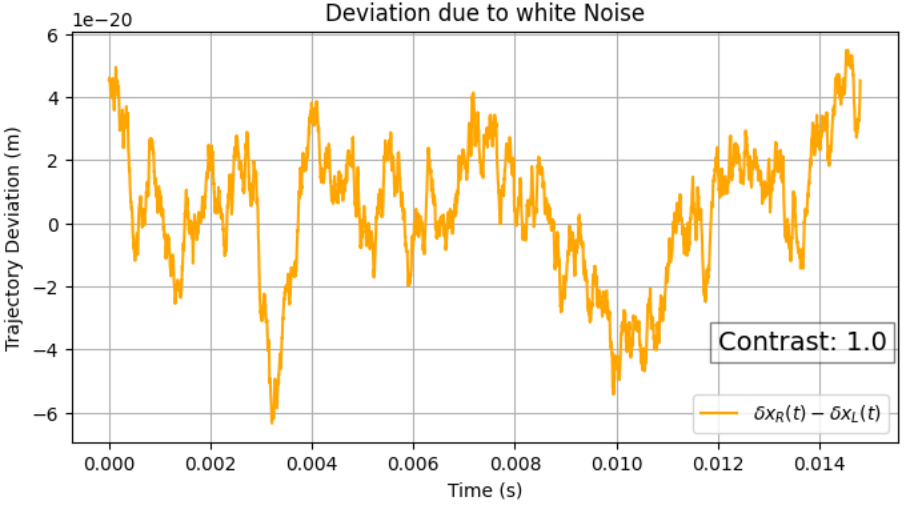}
        \caption{White noise}
        \label{fig:WhiteNoise_TrajectoryDeviation}
    \end{subfigure}
    \hfill
    \begin{subfigure}[b]{\linewidth}
        \includegraphics[width=\linewidth]{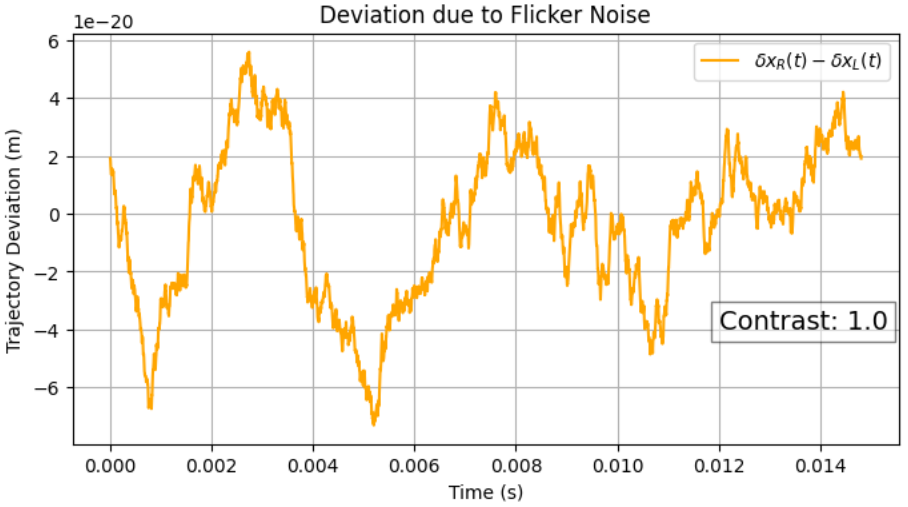}
        \caption{Flicker noise}
        \label{fig:FlickerNoise_TrajectoryDeviation}
    \end{subfigure}
    \captionsetup{justification=RaggedRight}
    \caption{
    Numerical simulations of trajectory deviations \( \delta x_R(t) - \delta x_L(t) \) for (a) white noise and (b) flicker noise, using system parameters from subsection~\ref{sec.a}. The $y$-axis is represented in terms of meters. The noise amplitudes are set according to bounds in eq.~(\ref{eq.WN_Bound}) and eq.~(\ref{eq.FN_Bound}), corresponding to a dephasing rate of \( \sim 10^2 \) Hz, respectively. In both cases, the resulting deviations are on the order of \( 10^{-19} \) m, which is several orders of magnitude smaller than the interferometer's spatial superposition length (\( \sim 10^{-9} \) m). Consequently, the contrast remains unity, indicating that noise-induced decoherence is negligible under these conditions.
    }\label{fig:TrajectoryDeviationComparison}
\end{figure}

From Fig.\ref{fig:WhiteNoise_TrajectoryDeviation} and Fig.\ref{fig:FlickerNoise_TrajectoryDeviation}, we observe that the contrast in the cases of white noise or flicker noise affecting the system is one. The deviations in the trajectory are extremely small ($\order{10^{-19}m}$) as compared to the size of the spatial superposition $\Delta x\sim {\cal O}({10^{-9}m})$. The plots have been generated for the values of the source parameters obtained in case of $\sim$10\% coherence (eq.(\ref{eq.WN_Bound}) for white noise and eq.(\ref{eq.FN_Bound}) for flicker noise). Hence the ensemble averages in each of the cases too will be expected to have high contrast.

The following is the explicit expression of the ensemble average of contrast in the case of white noise.
\begin{align}
    \langle\Delta x^2(T)\rangle &= \langle(\delta x_R(T) - \delta x_L(T))^2\rangle\nonumber\\
    &= \bigg(\frac{2\hbar\gamma_eA}{m\omega_0}\bigg)^2 \frac{2\pi}{\omega_0} = 4\times10^{-70}\\
    \langle\Delta p^2(T)\rangle &\approx \langle(\frac{2\hbar\gamma_e \eta_0}{\omega_0} )^2\rangle = 5.25\times10^{-22} \label{eq.wn_HD}
\end{align}
Thus, $\mathcal{C}(T) \approx 1$. This indicates that the noise in a magnetic field can be controlled such that the loss in contrast is negligible. The integral in the case of the flicker noise is more complex involving trigonometric integrals. However, analytical approximations in both the low- and high-frequency regimes yield results of comparable magnitude to eq.~(\ref{eq.wn_HD}). Thus, we conclude that loss of contrast due to Gaussian noise in current fluctuations bounded by values corresponding to a dephasing rate of 1$0^2$ Hz is negligible.

\section{Conclusions}
In this work, we have analyzed the impact of magnetic field gradient noise on the phase evolution of a quantum system modelled as a harmonic oscillator in an interferometric setup. By introducing noise at the Lagrangian level as a stochastic perturbation, we derived analytic expressions for the phase noise variance under two distinct spectral regimes: Gaussian white noise and flicker noise (1/f noise), both of which are relevant to real-world experimental environments. We provide a general expression for the effective transfer function of a harmonic oscillator subject to noise, which can be readily adapted to other physical systems with similar dynamics.

We established practical constraints on the noise amplitude for both noise types. For white noise, the gradient noise amplitude must satisfy: \linebreak \( A \leq 2.9 \times 10^{-6} \, \text{T} \, \text{m}^{-1} \text{Hz}^{-1/2} \), which corresponds to a maximum allowable current noise-to-signal ratio of \( \delta I/I \leq 1.3 \times 10^{-8} \). Similarly, for flicker noise, the source-dependent term was determined to be \( K \leq 0.7 \times 10^{-13} \,\text{T\,m$^{-1}$A$^{-1}$} \), leading to a comparable constraint of \( \delta I/I \leq 1.3 \times 10^{-8} \). These limits were derived using analytical and numerical methods, for system parameters such as the effective harmonic frequency \( \omega_0 \) derived following models for the Stern-Gerlach-type setup~\cite{Pedernales20_GM,Marshman:2021wyk}.

Thus, our theoretical analysis revealed that white noise leads to a higher dephasing rate compared to flicker noise. Nevertheless, both types of noise were found to impose nearly identical upper bounds on the permissible noise-to-signal ratio in the current source for a tolerable dephasing rate of approximately \( \Gamma \sim 10^2 \) Hz. This observation suggests that the experimental tolerance to current fluctuations is largely insensitive to the exact spectral character of the noise within a physically relevant range of the spectral exponent \( \alpha \in [0, 1.5] \). Such robustness provides a valuable degree of flexibility in experimental design and noise mitigation strategies.

Next, we turned to the Humpty-Dumpty problem of the fluctuations causing deviations in the trajectory such that the interferometer may not close. We developed a numerical simulation framework that generates synthetic noise in the frequency domain to compute the contrast of the interferometer output for a single run of the experiment.

The simulation results confirm that the trajectory deviations are several orders of magnitude smaller than the characteristic experimental superposition length scale (i.e., \( \sim 10^{-19} \, \text{m} \) compared to \( \sim 10^{-9} \, \text{m} \)). Consequently, the resulting contrast remains close to unity in both noise scenarios, indicating that the Humpty-Dumpty problem is negligible and that meaningful measurements can be conducted. Our findings deliver a theoretical and computational framework for assessing the effects of noise in magnetic field gradient in the Stern Gerlach-type interferometry.



\begin{acknowledgments}
AG is supported in part by NSF grants PHY-
2110524 and PHY-2111544, the Heising-Simons Foundation, the W.M. Keck Foundation, the John Templeton Foundation, DARPA, and
ONR Grant N00014-18-1-2370. 
S.B.'s work is supported by an EPSRC grant
EP/Y004590/1 MACON-QC.  
A.G. S.B. and A.M.'s research is funded by the Gordon and Betty Moore Foundation through Grant GBMF12328, DOI 10.37807/GBMF12328. This material is based on work supported by the Alfred P. Sloan Foundation under Grant No. G-2023-21130. SNM would like to thank the organizers of the workshop, Schrodinger Cats: The Quest to find the edge of the quantum world, held at OIST, Japan for facilitating academic interactions that led to this collaboration. SNM acknowledges support from the Kishore Vaigyanik Protsahan Yojana (KVPY) fellowship, SX-2011055, awarded by the Department of Science and Technology, Government of India.
S. B. would like to acknowledge EPSRC Grants No. EP/ N031105/1, No. EP/S000267/1, and No. EP/X009467/1 and STFC Grant No. ST/W006227/1

\end{acknowledgments}

\bibliography{References}
\begin{appendix}
\appendix
\section{Derivation of general form of the dephasing rate}
\label{appendix:A}

In this section, we derive the general form of the dephasing rate from the variation in the Lagrangian of a system due to noise. We ignore deviations in the trajectory. However, the method can be extrapolated to the same. We start with expressing the explicit time-dependence of the noisy terms in eq.(\ref{eq.genphasedif}), 
\begin{align}
    (\delta \phi \delta \phi^*)\quad\quad\quad\quad&\nonumber\\
    =\frac{1}{\hbar^2} \int_{t_i}^{t_f} \int_{t_i}^{t_f} \Bigg|\bigg[&(\delta A_R(t)x_R^2-\delta A_L(t)x_L^2) + (\delta B_R(t) x_R\nonumber\\
    &-\delta B_L(t) x_L) + (\delta C_R(t)-\delta C_L(t))\bigg]\,dt\Bigg|^2 
\end{align}
Generally, we know the statistics of the noise as a function of the frequency of the noise. Hence, we take a Fourier Transform of the time-dependent coefficients.
\begin{align}
    (\delta \phi \delta \phi^*) \quad\quad\quad \quad\quad&\nonumber\\
    = \frac{1}{\hbar^2} \Bigg|\int_{t_i}^{t_f} \int_{\omega_{min}}^{\infty}\bigg[&(\delta \tilde{A}_R(\omega)x_R^2-\delta \tilde{A}_L(\omega)x_L^2) \nonumber\\
    &+ (\delta \tilde{B}_R(\omega) x_R -\delta \tilde{B}_L(\omega) x_L)\nonumber\\
    & + (\delta \tilde{C}_R(\omega)-\delta \tilde{C}_L(\omega))\bigg] e^{i\omega t}\, d\omega\,dt\Bigg|^2 
\end{align}
Now, we invoke the assumption that $\delta A_L$ and $\delta B_L$ are some linear functions in $\delta A_R$ and $\delta B_R$ respectively. 
Let,
\begin{align}
    \delta A_j(t) &= S_{xj}D_{As}f_{As}(t) + D_{An}f_{An}(t)\\
    \delta B_j(t) &= S_{xj}D_{Bs}f_{Bs}(t) + D_{Bn}f_{Bn}(t)\\
    \delta C_j(t) &= D_{Cj}f_{Cj}(t)
\end{align}
where $D_{An}$ is the coefficient of the spin-independent part of $\delta A_j$ that is dependent on noise.
\begin{align}
    \delta \phi \delta \phi^*& \nonumber\\
    =\frac{1}{\hbar^2}  &\Bigg|\int_{t_i}^{t_f} \int_{\omega_{min}}^{\infty} \Big[\tilde{f}_{An}(\omega)D_{An}(x_R^2-x_L^2) \nonumber\\
    &+ \tilde{f}_{As}(\omega)D_{As}(x_R^2+x_L^2)+ \tilde{f}_{Bn}(\omega)D_{Bn}(x_R-x_L) \nonumber\\
    &+ \tilde{f}_{Bs}(\omega)D_{Bs}(x_R+x_L) + D_{CR}\tilde{f}_{CR}(\omega)\nonumber\\
    &-D_{CL}\tilde{f}_{CL}(\omega) \Big] \times e^{i\omega t} \, d\omega\,dt \Bigg|^2
\end{align}
Now consider a single source of noise. Hence,
\begin{align}
    f_{As}(t) = f_{An}(t) = f_{Bs}(t) = f_{Bn}(t) = f_{Cj}(t) = f(t)
\end{align}
Hence,
\begin{align}
    \delta \phi \delta \phi^*& \nonumber\\
    = \frac{1}{\hbar^2} &\Bigg|\int_{t_i}^{t_f} \int_{\omega_{min}}^{\infty}\tilde{f}(\omega)\bigg[ D_{An}(x_R^2-x_L^2) + D_{As}(x_R^2+x_L^2) \nonumber\\
    &+ D_{Bn}(x_R-x_L) + D_{Bs}(x_R+x_L) \nonumber\\
    &+ (D_{CR}-D_{CL})\bigg] e^{i\omega t}\, d\omega\,dt\Bigg|^2 \\
    = \frac{1}{\hbar^2} &\int_{\omega_{min}}^{\infty}\int_{\omega_{min}}^{\infty}\tilde{f}(\omega) \tilde{f}(\omega') \nonumber\\
    &\Bigg|\int_{t_i}^{t_f} \bigg[ D_{An}(x_R^2-x_L^2) + D_{As}(x_R^2+x_L^2) \nonumber\\
    &+ D_{Bn}(x_R-x_L) + D_{Bs}(x_R+x_L) \nonumber\\
    &+ (D_{CR}-D_{CL})\bigg] e^{i\frac{(\omega+\omega')}{2} t}\,dt\Bigg|^2 \, d\omega\, d\omega'
\end{align}
Ultimately, what concerns us is the variance of the phase difference due to noise:
\begin{align}
    E[\delta \phi \delta \phi^*]&\nonumber\\
    = \frac{1}{\hbar^2}& \int_{\omega_{min}}^{\infty}\int_{\omega_{min}}^{\infty}E[\tilde{f}(\omega) \tilde{f}(\omega')]\nonumber\\
    &\Bigg|\int_{t_i}^{t_f} \bigg[ D_{An}(x_R^2-x_L^2) + D_{As}(x_R^2+x_L^2) \nonumber\\
    &+ D_{Bn}(x_R-x_L) + D_{Bs}(x_R+x_L) \nonumber\\
    &+ (D_{CR}-D_{CL})\bigg] e^{i\frac{(\omega+\omega')}{2} t}\,dt\Bigg|^2 \, d\omega\, d\omega'
\end{align}
Thus, from eq.(\ref{eq.Gammaandensemble}), we obtain:
\begin{align}
    \Gamma
    = \frac{1}{\hbar^2} &\int_{\omega_{min}}^{\infty} S(\omega) \Bigg|\int_{t_i}^{t_f} \bigg[ D_{An}(x_R^2-x_L^2) + D_{As}(x_R^2+x_L^2) \nonumber\\
    &+ D_{Bn}(x_R-x_L) + D_{Bs}(x_R+x_L) \nonumber\\
    &+ (D_{CR}-D_{CL})\bigg] e^{i\omega t}\,dt\Bigg|^2 \, d\omega\\
    = \frac{1}{\hbar^2}& \int_{\omega_{min}}^{\infty} S(\omega) \bigg|\Sigma_i\sqrt{F_i(\omega)}\bigg|^2 \, d\omega    
\end{align}
where $F_1(\omega) = \left| \int_{t_i}^{t_f} dt\, D_{An}(x_R - x_L) e^{i \omega t} \right|^2$.

\section{Analytical expression for dephasing rate in case of white noise}
\label{appendix:B}
\begin{figure}
    \centering
    \includegraphics[width=\linewidth]{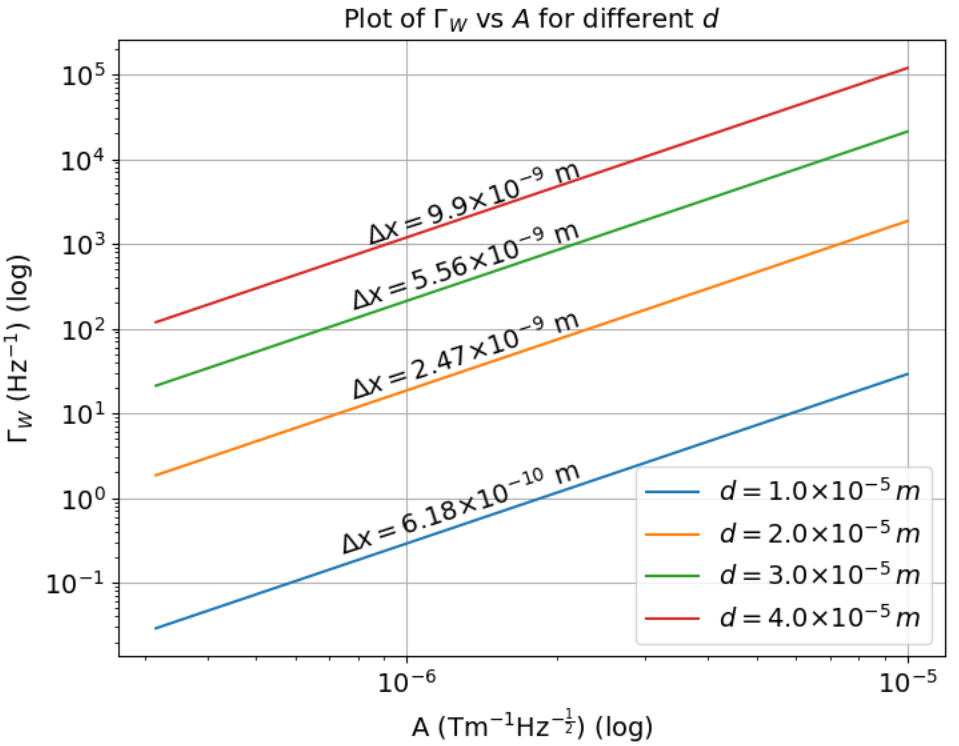}
    \captionsetup{justification=RaggedRight}
    \caption{Log-Log plot of \( \Gamma_W \) versus \( A \) for various values of the separation distance \( d \). The data exhibit a clear linear trend, indicating a power-law relationship. Each curve corresponds to a different distance \( d \) of the particle from the wire, with associated maximum superposition distance being \( \Delta x \) labeled. The slope remains consistent across curves, while the intercept varies with \( d \), highlighting its influence on noise sensitivity. This is the log-log plot of the Fig.(\ref{fig:WN for dif d gvA}). }\label{fig:WN_for_dif_d_gvA_Gamma_102}
\end{figure}

In Fig.(\ref{fig:WN_for_dif_d_gvA_Gamma_102}), we observe that the log-log plot of $\Gamma_{W}$ vs ${A}$ follows a linear trend. Fitting the graphs, we obtain:
\begin{align}
    \log(\Gamma_{W}) &\approx 2\log(A)+ c_W(d)
\end{align}
where $c_W(d)$ is the y-intercept as a function of d. We obtain the following relation for $c_W(d)$ via graphical analysis:
\begin{equation}
    c_W(d) \approx 6\log(d)+41.4680
\end{equation}
For d = $2\times10^{-5} $m to a linear curve, we obtain: 
\begin{align}
    \log(\Gamma_{W}) &\approx 2\log(A)+ 13.274\nonumber\\
    \implies \Gamma_{W} &\approx 1.88A^2\times10^{13}
\end{align}

\section{Analytical expression for dephasing rate in case of flicker noise}
\label{appendix:C}


Linear fitting of the graphs in Fig.(\ref{fig:flickernoise}) yields:
\begin{align}
    \log(\Gamma_{F}) &\approx \log(K) + c_F(d)
\end{align}
where $c_F(d)$ is the y-intercept as a function of d. We obtain the following relation for $c_F(d)$ via graphical analysis:
\begin{equation}
    c_F(d) \approx 6\log(d)+43.5567
\end{equation}
Thus, corresponding to $d=20\mu $m we obtain:
\begin{align}
    \log(\Gamma_{F}) &\approx \log(K) + 15.3629\nonumber\\
    \implies \Gamma_{F} &\approx 2.31K \times 10^{15}
\end{align}
\begin{figure}[ht!]
    \centering
    \includegraphics[width=\linewidth]{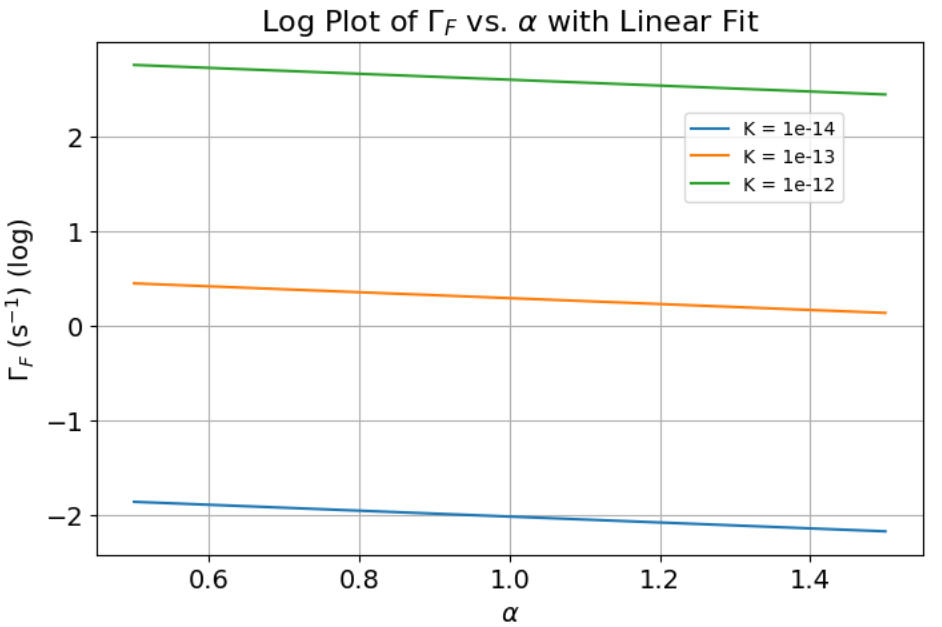}
    \captionsetup{justification=RaggedRight}
    \caption{Plot of \( \Gamma_F \) versus spectral exponent \( \alpha \) for noise strengths \( K = 10^{-14}, 10^{-13}, 10^{-12} \) in the case of flicker(1/f) noise. While the main analysis in the paper focuses on \( \alpha = 1 \), this plot demonstrates that the dephasing rate \( \Gamma_F \) remains within the same order of magnitude across a range of \( \alpha \) values for each fixed \( K \). This supports the validity of extrapolating the constraints derived at \( \alpha = 1 \) to systems characterized by other values of \( \alpha \).}  \label{fig:flickernoise_multiK}
\end{figure}
To assess the robustness of our results with respect to variations in the spectral exponent \( \alpha \), we compute the dephasing rate \( \Gamma_F \) as a function of \( \alpha \) for different values of the noise strength \( K \). While our primary analysis focuses on the case \( \alpha = 1 \), corresponding to 1/f noise, the results shown in Fig.(\ref{fig:flickernoise_multiK}), demonstrate that \( \Gamma_F \) remains within the same order of magnitude across a range of \( \alpha \) values for each fixed \( K \). This indicates that the scaling of the dephasing rate is not highly sensitive to the precise value of \( \alpha \), thereby justifying the extension of our constraints on $\frac{\delta I}{I}$ to materials or environments where \( \alpha \neq 1 \).


\section{Dephasing rate on inclusion of the deviation in trajectory due to noise}
\label{appendix:D}

On substituting for deviations in the trajectory (eq.(\ref{eq.dev_traj_fouriersoln})) in eq.(\ref{eq.pert_action_complete}), we obtain the total phase difference between the two arms of the interferometer due to noise:
\begin{align}
    \delta \phi 
    =H&\bigg(\int_{-\infty}^{-\omega_{min}}+\int_{\omega_{min}}^{\infty}\bigg)\Bigg[\delta\tilde{\eta}(\omega)\nonumber\\
    &\int_0^T\bigg\{ \frac{\cos(\omega_0 t)(\cos(\omega_0 t) - 1) }{\omega_0^2}
     \nonumber\\
    &+ \frac{(2\cos(\omega_0t) - 1)\cos(\omega_0 t)}{(\omega^2-\omega_0^2)}  \,\bigg\}e^{i\omega t}\,dt\Bigg]\,d\omega \label{eq.total_phasedif}
\end{align}
Dephasing rate accounting for terms arising only from deviations in the trajectory arises from the second term in eq.(\ref{eq.total_phasedif}). This is given by :
\begin{align}
    \Gamma_{\text{dev}} &= \frac{8H^2}{\omega_0^5} \int_{1}^{\infty} S(\xi)F_{\text{dev}}(\xi) d\xi \label{eq.G_dev}
\end{align}
where, $\xi=\frac{\omega}{\omega_0}$
\begin{equation}
    F_{\text{dev}}(\xi) =  \frac{\sin^{2}\left(\pi \xi\right)}{(1-\xi^2)^2} \bigg[ \dfrac{1}{\xi} - \frac{\xi}{\xi^2-1} +\frac{\xi}{\xi^2-4} \bigg]^2
\end{equation}
Thus, from eq.(\ref{eq.Dephasing_genform}), we obtain the following to be the additional contribution to the dephasing rate:
\begin{align}
    \Gamma_{tot} &= \frac{8H^2}{\omega_0^5} \int_{1}^{\infty}S(\xi)\bigg(\sqrt{F_{HO}(\xi)}+\sqrt{F_{\text{dev}}(\xi)}\bigg)^2 d\xi \label{eq.G_tot}
\end{align}
We perform an order-of-magnitude estimate using parameter values drawn from Sec.~\ref{sec.a} and Sec.~\ref{sec.5} to verify that the dephasing rate due to trajectory deviations, denoted \( \Gamma_{\text{dev}} \), does not exceed the order of magnitude of dephasing rate calculated without accounting for such deviations (i.e., Eq.~(\ref{eq.G_d})). This estimation is carried out using using values in Table.(\ref{tab:estimate}).
\begin{table}[!ht]
\centering
\begin{tabular}{|l|l|}
\hline
\textbf{Quantity} & \textbf{Value} \\
\hline
Typical position, \( x_j \) & \( 10^{-9} \, \mathrm{m} \) \\
\hline
Magnetic field gradient, \( \eta_0 \) & \( 6 \times 10^3 \, \mathrm{T/m} \) \\
\hline
Trajectory deviation, \( \delta x_j \) & \( 10^{-15} \, \mathrm{m} \) \\
\hline
Gradient fluctuation, \( \delta \eta = \frac{\delta I}{I}\eta_0 \) & \( 6 \times 10^{-5} \, \mathrm{T/m} \) \\
\hline
\end{tabular}
\captionsetup{justification=RaggedRight}
\caption{Order-of-magnitude values used in estimating dephasing contributions. The values of $x_j$, $\eta_0$ and $\frac{\delta I}{I}$ are consistent with values used throughout the paper. The value of \( \delta x_j \) is obtained from numerical simulations similar to those in Sec.~\ref{sec.5}.}
\end{table}\label{tab:estimate}
Using these values, we estimate the two types of contributions to the dephasing in eq.(\ref{eq.pert_action_complete}), one involving the fluctuations directly in the magnetic field gradient and the other involving the fluctuations in the trajectory as result of fluctuations in the field gradient. We note from eq.(\ref{eq.pert_action_complete}) that the dephasing is proportional $x_j \delta \eta$ and $\delta x_j \eta_0$. Hence we use these values to estimate the order of magnitude of the two types of contributions to the dephasing.
\begin{align}
x_j \delta \eta &\sim 10^{-13} \, \mathrm{T} \label{eq.det_deph}\\
\delta x_j \eta_0 &\sim 10^{-11} \, \mathrm{T} \label{eq.dev_deph}
\end{align}
These contributions suggest that the dephasing due to deviations cannot be neglected. Hence, we have to explicitly account for the contribution of the deviations in the trajectory to the dephasing.


From the above analysis we also note that though the deviations in trajectory may be small enough to close the interferometer, i.e. we can have a contrast = 1, the deviations in the trajectory contribute to the dephasing and has to be accounted for while obtaining precise bounds on experimental parameters. 

\section{Nature of the Transfer Function and PSD}
\label{appendix:E}
\begin{figure}[!ht]
    \centering
    \begin{subfigure}[b]{\linewidth}
        \includegraphics[width=\linewidth]{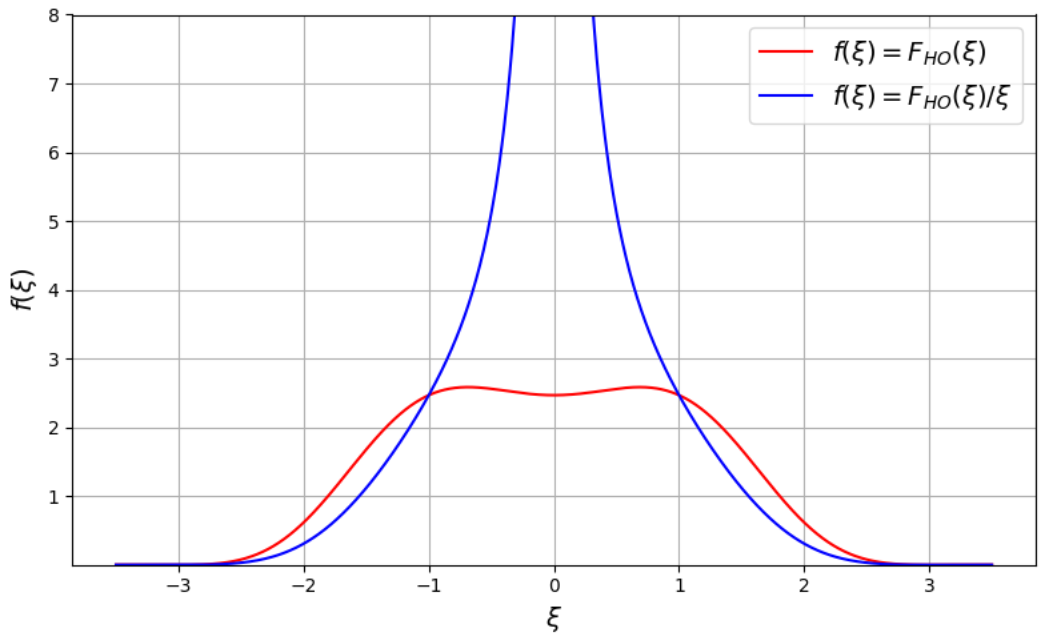}
        \label{fig:F_HO_paper_notlog}
    \end{subfigure}
    \caption{
    The graph show the nature of the integrands in eq.(\ref{eq.G_d}) as a function of $\xi = \frac{\omega}{\omega_0}$ in case of white noise (red curve) and flicker noise (blue curve), respectively. 
    }\label{fig:F_HO paper}
\end{figure}

To gain insight into how different noise spectra contribute to dephasing, we examine the behavior of the integrand in Eq.\eqref{eq.G_d}, plotted in Fig.(\ref{fig:F_HO paper}) as a function of the normalized frequency \( \xi=\omega/\omega_0 \). The red curve corresponds to white noise, characterized by a flat spectral density, while the blue curve represents flicker noise, where the spectral density scales as \( 1/\omega \). For \( \xi>1 \), the integrand associated with flicker noise lies below that of white noise, reflecting the fact that flicker noise is predominantly concentrated at low frequencies. Despite the constant strength of white noise across the spectrum, the system’s frequency response \( F_{HO}(\omega) \) suppresses contributions from higher frequencies, ensuring that the dephasing rate remains finite. The figure also highlights that the apparent singularities in the harmonic oscillator response function at \( \xi=1 \) and \( \xi=2 \), as suggested by Eq.~\eqref{eq.FHO_transferfnt}, are in fact removable and do not lead to divergences. These highlight how the interplay between the noise spectrum and the system’s susceptibility determines the extent of dephasing, and they help justify the convergence properties of the integral in both the white and flicker noise cases. 

\end{appendix}
\end{document}